\documentclass[journal,12pt,onecolumn,draftclsnofoot]{IEEEtran}
\IEEEoverridecommandlockouts
\usepackage{cite}
\usepackage{amsmath,amssymb,amsfonts}
\usepackage{algorithmic}
\usepackage{graphicx}
\usepackage{textcomp}
\usepackage{xcolor}
\usepackage{dsfont}
\usepackage{subcaption}
\usepackage{float}
\usepackage{amsmath}
\usepackage[]{graphicx}
\usepackage{lipsum}
\usepackage{stackengine}
\usepackage{url}%add this always
\usepackage{hyperref}%add this always
\usepackage{algorithmic}

\def\BibTeX{{\rm B\kern-.05em{\sc i\kern-.025em b}\kern-.08em
    T\kern-.1667em\lower.7ex\hbox{E}\kern-.125emX}}

\begin{document}

\title{Non-Orthogonal Multiplexing of Ultra-Reliable and Broadband Services in Fog-Radio Architectures\\
\thanks{The work of Rahif Kassab, Osvaldo Simeone and Petar Popovski has received funding from the European  Research  Council (ERC) under the European Union Horizon 2020 research and innovation program (grant agreements 725731 and 648382). The work of O. Simeone has also been supported by U.S. NSF under grant  CCF-1525629.}
}

\author{\IEEEauthorblockN{Rahif Kassab\IEEEauthorrefmark{1},
Osvaldo Simeone\IEEEauthorrefmark{1}, Petar Popovski\IEEEauthorrefmark{2} and Toufiqul Islam\IEEEauthorrefmark{3} }\\
\IEEEauthorblockA{\small\IEEEauthorrefmark{1}Centre for Telecommunications Research, King's College London, London, United Kingdom\\
\IEEEauthorrefmark{2}Department of Electronic Systems, Aalborg University, Aalborg, Denmark\\
\IEEEauthorrefmark{3} Intel Corporation, Santa Clara, CA, USA \\
Emails: \IEEEauthorrefmark{1}\{rahif.kassab,osvaldo.simeone\}@kcl.ac.uk,
\IEEEauthorrefmark{2}petarp@es.aau.dk},\IEEEauthorrefmark{3}toufiqul.islam@intel.com}
\vspace{-10 mm}
\maketitle

\begin{abstract}
The fifth generation (5G) of cellular systems is introducing Ultra-Reliable Low-Latency Communications (URLLC) services alongside more conventional enhanced Mobile BroadBand (eMBB) traffic. Furthermore, the 5G cellular architecture is evolving from a base station-centric deployment to a fog-like set-up that accommodates a flexible functional split between cloud and edge. In this paper, a novel solution is proposed that enables the non-orthogonal coexistence of URLLC and eMBB services by processing URLLC traffic at the Edge Nodes (ENs), while eMBB communications are handled centrally at
a cloud processor as in a Cloud-Radio Access Network (C-RAN) system. This solution guarantees the low-latency requirements
of the URLLC service by means of edge processing, e.g., for vehicle-to-cellular use cases, as well
as the high spectral efficiency for eMBB traffic via centralized baseband processing. Both uplink and downlink are analyzed by accounting for the heterogeneous performance requirements of eMBB and URLLC traffic and by considering practical
aspects such as fading, lack of channel state information for URLLC transmitters, rate adaptation for
eMBB transmitters, finite fronthaul capacity, and different coexistence strategies, such as puncturing. 
\end{abstract}


\section{Introduction}
\label{sec:introduction}
With the advent of 5G, wireless cellular systems are undergoing an evolution in terms of both services and network architecture. Conventional enhanced Mobile BroadBand (eMBB) services, mostly aimed at consumers, will share radio and network resources with Ultra-Reliable Low-Latency Communications (URLLC) and machine-type traffic, which cater to vertical industries \cite{petar_urllc_review}. Furthermore, the supporting cellular network architecture will evolve from a traditional base station-centric deployment to a fog-like set-up \cite{3gpp_ran}\cite{5g2016view} with computation and communication resources at both edge and cloud. Thanks to network softwarization, this architecture will enable network functionalities to be distributed among edge and cloud elements\footnote{These are defined as distributed and central units by 3GPP \cite{3gpp_ran}, respectively.} depending on their latency and reliability requirements \cite{3gpp_ran,5g2016view,tandon2016harnessing}. An extreme instance of this type of network architectures is Cloud-Radio Access Network (C-RAN), in which all processing, apart from radio-frequency components, is carried out in the cloud \cite{bookcransimeone}.\par

\begin{figure}[t]
	\centering
	\includegraphics[height= 6  cm, width= 6.5 cm]{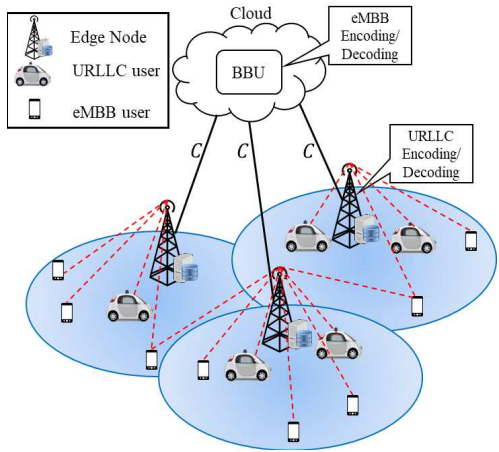}
	\caption{An F-RAN multi-cell system with coexisting eMBB and URLLC.}
	\label{fig:system_model}
\end{figure}

The coexistence of the heterogeneous services eMBB and URLLC is typically envisioned to rely on the orthogonal allocation of spectral resources to the two traffic types. Orthogonal multiplexing ensures the isolation of eMBB and URLLC traffic. This, in turn, enables the application of  ``slicing'', a Radio Resource Management approach introduced in 5G that carries out separate designs to meet the heterogeneous performance metrics and guarantees of the two services. As an alternative to orthogonal resource allocation, Non-Orthogonal Multiple Access (NOMA) is grounded by classical results in information theory that prove the capacity-achieving property of non-orthogonal transmissions in multiple access channels and of superposition coding in broadcast channels, modeling single-cell Uplink (UL) and Downlink (DL) scenarios, respectively \cite{cover2012elements}. Motivated by these information-theoretic results, NOMA has been proposed as a key component of 5G systems as a mean to share radio resources among transmissions belonging to \textit{homogeneous} devices \cite{downlink_noma_shortpacket,ding2016designnomaULDL,ding2018powerallocation}. A first question that motivates this work is: \textit{Can NOMA improve the performance of the coexistence between the heterogenuous eMBB and URLLC services?} \par
An illustration of a fog-based multi-cell system, also known as Fog-RAN (F-RAN)\cite{tandon2016harnessing}\cite{kang2018control}\cite{peng2016fog} is provided in Fig. \ref{fig:system_model}. In this system, each cell contains an Edge Node (EN), along with its connected computing platform, as well as multiple eMBB and URLLC users. All ENs have a finite-capacity digital fronthaul link to a cloud processor. The mentioned flexibility in the allocation of functions between ENs and cloud afforded by network softwarization motivates the second question that guides this work: \textit{How can the physical layer network functionalities be split between edge and cloud in order to improve the coexistence of eMBB and URLLC traffic?}
%%%%%%%%%%%%%%%%%%%%%%%%%%%%%%%%%%%%%%%%%%%%%%%%%%%%%%%%%%%%%%%%%%%%%%%%%%%%%%%%%%%%%%%%%%%%%%%%%%%%
\subsection{Main Contributions} 
In this work, we address the first question by considering the performance of both eMBB and URLLC traffic in the F-RAN multi-cell system of Fig. \ref{fig:system_model} under both orthogonal and non-orthogonal multiple access.
Following \cite{networkslicingpopvskiandsimeone}, we refer to this latter approach as \textit{Heterogeneous NOMA} (H-NOMA), in order to highlight the key distinction with respect to conventional NOMA of accommodating services with heterogeneous, rather than homogeneous, performance requirements. \par
Moreover, we tackle the second question by proposing a novel cloud-edge functional split for the coexistence of URLLC and eMBB traffic. Accordingly, URLLC traffic is handled at the ENs in order to meet the low-latency requirements, URLLC devices may be vehicles in vehicle-to-cellular use cases\cite{link_mit_vehicles}, or they may be devices serving automation chains in Industry 4.0 scenarios characterized by automation and communication-based manufacturing \cite{slicing_industry}\cite{indstry4requirements}. In contrast, eMBB traffic is processed by the centralized BaseBand Unit (BBU) in the cloud as in a C-RAN architecture in order to enhance spectral efficiency thanks to the cloud's interference management capabilities \cite{boviz2017effective}. The proposed hybrid edge-cloud solution fully leverages the unique features of F-RAN systems in order to cater to the heterogeneous requirements of URLLC and eMBB systems.\par
% As a brief review, NOMA of eMBB transmissions is considered in \cite{ding2016designnomaULDL} for the UL and the DL of a single cell system with a multiple-antenna base station. In \cite{ding2016designnomaULDL}, a transmission scheme based on NOMA was proposed for UL and DL based on signal alignment and on different power allocation strategies. Reference \cite{ding2018powerallocation} considers the power allocation problem in NOMA for the UL where the goal is to minimize the energy consumption given delay constraints. NOMA was considered for URLLC in \cite{downlink_noma_shortpacket} for a two-user DL broadcast channel, in which the effective throughput of one user is maximized while guaranteeing requirements on the effective throughput of the other user. NOMA for DL C-RAN was studied in \cite{nomacranDL} using stochastic geometry, and closed-form expressions for the outage probability of users at cell edges were derived. Finally, in \cite{outage_naofal} the effect of joint transmitter/receiver in-phase and quadrature-phase imbalance on the performance of NOMA was analyzed for the DL.\par
\par We analyze and compare the performance of conventional Heterogeneous-Orthogonal Multiple Access (H-OMA) techniques with H-NOMA in the \textit{multi-cell F-RAN} system of Fig.~\ref{fig:system_model} for both Uplink (UL) and Downlink (DL). As illustrated in Fig.~\ref{fig:time_frequency}, the communication model assumes random activation of URLLC users, which implies possible collisions in the UL and blockages in the DL, and scheduled access for eMBB users. We aim at understanding the possible tradeoffs between the eMBB and URLLC performance as a function of key parameters, such as the capacity of the digital fronthaul links connecting cloud and ENs, as well as the deployed strategy including puncturing, treating URLLC interference as noise, and superposition coding (see Fig.~\ref{fig:system_model_noma_UL} for the UL and Fig.~\ref{fig:system_model_noma_DL} for the DL). For a preview of the main results, we refer to Fig. \ref{fig:UL_function_a_U}, which shows the per-cell eMBB and URLLC achievable rates as function of the URLLC traffic load for both H-OMA and H-NOMA.
%%%%%%%%%%%%%%%%%%%%%%%%%%%%%%%%%%%%%%%%%%%%%%%%%%%%%%%%%%%%%%%%%%%%%%%%%%%%%%%%%%%%%%%%%%%%%%%%%
\subsection{Related Work}
%%%%%%NOMA UL and DL
In the past few years, NOMA has been widely investigated as a solution to increase the spectral efficiency of cellular networks for both UL and DL. The key idea is to allow simultaneous transmissions on the UL and superposition coding on the DL. As some representative examples, in the UL case, reference \cite{ul_noma_2} shows that NOMA with Successive Interference Cancellation (SIC) at the base stations can significantly enhance cell-edge users' throughput, while paper \cite{ul_noma_1} proposes a NOMA scheme based on joint processing at the base stations. As for the DL, NOMA was demonstrated to achieve superior performance in terms of ergodic sum rate of a cellular network with randomly deployed users in\cite{ding2014performance}. Furthermore, in reference \cite{saito2013system}, OMA techniques were compared with NOMA under SIC in terms of system-level performance by taking into account key LTE functionalities such as Hybrid Automatic Repeat reQuest (HARQ) and time/frequency domain scheduling. Other related works include \cite{noma_other_2},\cite{noma_other_3} and \cite{noma_other_4}.\par
%%%%%%%H-OMA resource allocation and MAC
While all the work discussed above assumes homogeneous services, coexistence between heterogeneous services such as eMBB and URLLC has been mostly studied under the assumption of orthogonal resource allocation. For example, in \cite{embb_urllc_oma} a null-space-based spatial preemptive scheduler for URLLC and eMBB traffic is proposed that aims at guaranteeing URLLC quality of service while maximizing the eMBB ergodic capacity. In \cite{embb_urllc_fog}, the joint user-base station association and orthogonal resource allocation problem was considered for the DL of a fog-network in the presence of eMBB and URLLC services.\par
%%%%%%%H-NOMA
The non-orthogonal coexistence of heterogeneous services has been much less studied. In \cite{Anand2017JointSO}, the joint scheduling of eMBB and URLLC has been investigated with the goal of maximizing the utility of eMBB traffic while satisfying the quality of service requirements of URLLC traffic. The problem formulation accounts for the different time scales of traffic generation of the two services. 
% In \cite{efficient_mmtc_urllc} a multi-armed bandit reinforcement learning approach is investigated to achieve the optimal tradeoff between feedback and feedbackless based transmission to support URLLC and mMTC.
From an information-theoretic standpoint, the coexistence of heterogeneous services was studied under the rubric of unequal error protection in simplified abstract settings in \cite{amos2017}. A communication-theoretic model for the coexistence of eMBB-URLLC and eMBB-mMTC was introduced in \cite{popovski20185g} for a \textit{single-cell} model with decoding at the base station. To the best of our knowledge, the multi-cell case was only studied by some of the authors in \cite{rahifuplink} by considering \textit{only the UL} and a simplified Wyner-type channel model with \textit{no fading and inter-cell interference limited to neighbouring cells}. We also refer to \cite{andrea_uplink_cran} that considers a setup with the same limitations as \cite{rahifuplink} but with analog fronthaul links. Another related theoretical work for the UL Wyner model is \cite{wiggerdelay}, in which higher-latency messages are decoded by means of cooperation between adjacent cells, while lower-latency messages are decoded without cooperation. 

%%%%%%%%%%%%%%%%%%%%%%%%%%%%%%%%%%%%%%%%%%%%%%%%%%%%%
\subsection{Organization of the paper}
The rest of the paper is organized as follows. In the next section, we detail the system model, while Sec. \hyperref[sec:signal_model]{III} describes the signal model and the performance metrics. In Sec. \hyperref[sec:UL_OMA]{IV} and \hyperref[sec:DL_OMA]{V}, the performance of H-OMA is evaluated for UL and DL respectively, while in Sec. \hyperref[sec:UL_NOMA]{VI} and \hyperref[sec:DL_NOMA]{VII}, the performance of H-NOMA is analyzed for UL and DL respectively. Finally, numerical results are presented in Sec. \hyperref[sec:numerical_results]{VIII}, and conclusions are drawn in Sec. \hyperref[sec:conclusion]{IX}.
\par \textbf{Notation:} Bold upper-case characters denote matrices and bold lower-case characters denote vectors. $\mathbb{E}_{X}[\cdot]$ represents the expectation of the argument with respect to the distribution of the random variable $X$. $\mathbf{A}^{\mathsf{H}}$ denotes the Hermitian transpose of matrix $\mathbf{A}$. $X \sim \mathcal{B}\mathbf{\textit{ern}}(p)$ indicates a Bernoulli distribution with parameter $p$. $Y \sim \mathcal{B}\mathbf{\textit{in}}(n,p)$ indicate a Binomial random variable distribution with parameters $n$ and $p$. $I_z(X;Y)$ denotes the mutual information between random variables $X$ and $Y$ for the given constant value $z$ of random variable $Z$, i.e., $I_z(X;Y)  = I(X;Y|Z=z)$. $|\mathbf{A}|$ is the determinant of matrix $\mathbf{A}$.

%%%%%%%%%%%%%%%%%%%%%%%%%%%%%%%%%%%%%%%%%%%%%%%%%%%%%%%%%%%%%%%%%%%%%%%%%%%%%%%%%%%%%%%%%%%%%
\section{System Model}
\label{sec:system_model}
As illustrated in Fig. \ref{fig:system_model}, we study UL and DL communications in a cellular network with an F-RAN architecture that encompasses both eMBB and URLLC users. Each one of the $M$ cells contains an Edge Node (EN) and multiple eMBB and URLLC users. All ENs are connected to a Baseband Unit (BBU) in the cloud by mean of orthogonal fronthaul links of capacity $C$ bit/s/Hz, or equivalently $C$ bits for each symbol of the wireless channel. The RAN uses Frequency Division Duplex (FDD) in order to facilitate grant-free URLLC transmissions, as detailed below.
\par \textit{F-RAN topology and operation:} As illustrated in Fig. \ref{fig:system_model}, we assume that the URLLC users are located close to the ENs, and hence URLLC communications take place with non-negligible power only with the EN in the same cell. As a result, URLLC users do not cause interference to ENs in other cells while transmitting in the UL, and they only receive transmissions from the same-cell EN in the DL. This condition can be ensured by allowing URLLC transmissions only from users with large average channel gain to the target EN, so that the high reliability requirement of URLLC traffic can be satisfied. As an example, as seen in Fig. 1, the EN may serve a nearby vehicle for transmission of time-sensitive control information in vehicle-to-cellular use cases \cite{link_mit_vehicles}. Alternatively, in mission-critical or Industry 4.0 scenarios, ENs can be deployed in locations that contain URLLC devices. The eMBB users, instead, need not guarantee this condition, and are assumed to be in arbitrary positions with potentially non-negligible channel gains to all ENs for both UL and DL. 
\par As illustrated in Fig.~\ref{fig:system_model}, for the UL, due to latency constraints, the URLLC signals are decoded locally at the EN, while the eMBB traffic is decoded centrally at the BBU as in a C-RAN architecture\cite{zhou2014optimized} \cite{park_robust}. In a similar manner, in the DL, the eMBB traffic is assumed to be generated at the cloud, e.g., as a result of web searches or broadband streaming, and C-RAN precoding and quantization are applied \cite{zhou2016fronthaul} \cite{park2013joint}. In contrast, URLLC traffic is generated at the edge, with each EN serving same-cell URLLC users, in line with the use cases mentioned above. Note that these assumptions imply that the higher layers of eMBB and URLLC services are implemented separately at cloud and edge, respectively.\par
\begin{figure}[t]
	\centering
	\includegraphics[height= 7 cm, width= 9 cm]{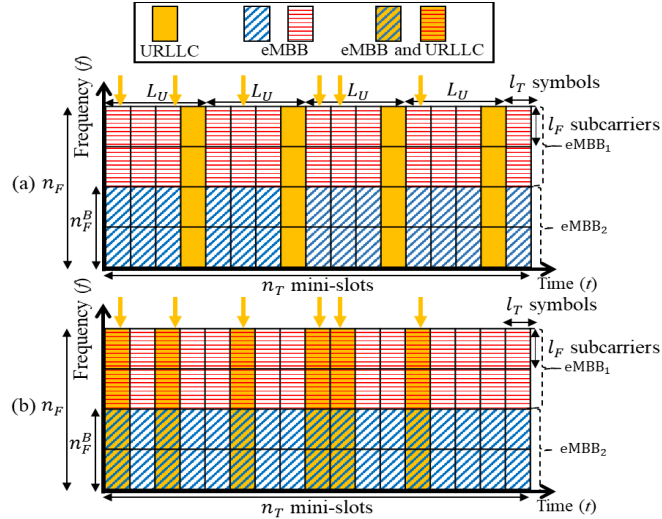}
	\caption{Time-frequency resource allocation for: (a) Heterogeneous-Orthogonal Multiple Access Scheme (H-OMA); and (b) Heterogeneous-Non Orthogonal Access Scheme (H-NOMA); Hatched areas correspond to eMBB transmissions. Downward arrows denote generation of URLLC packets to or from different URLLC users.}
    \vspace{-5 mm}
	\label{fig:time_frequency}
\end{figure}
\textit{Frame structure:} As illustrated in Fig.~\ref{fig:time_frequency}, we consider a radio interface that operates in frames of $n_T$ mini-slots and $n_F$ frequency channels. The time frequency plane is organized into Resource Units (RUs), and each RU spans one mini-slot of index $t \in \{1,\ldots,n_T \}$ and one frequency channel of index $f \in \{1,\ldots,n_F \}$, and it contains $l_T$ symbols in time domain and $l_F$ subcarriers\footnote{As an example, in 3GPP release 15 \cite{3gpp_release15}, the RU consists of $l_F$ =12 subcarriers and $l_T$=14 symbols.}. We index as $(f,t)$ the RU located at frequency channel $f$ and mini-slot $t$.
\par eMBB traffic is scheduled and a single eMBB user is assigned to all $n_T$ minislots in a frame and a set $n_F^B \leq n_F$ frequency channels, so that there are $\lfloor n_F/n_F^B \rfloor$ scheduled eMBB users per-cell. In contrast, URLLC transmissions are assumed to be grant-free, and packets are randomly generated in each mini-slot for URLLC users. As a result, the number of active URLLC users per-cell in each frame is random. Furthermore, due to latency constraints, each URLLC transmission can span only a single mini-slot, and hence the blocklength of an URLLC transmission is equal to $n_{F}l_Fl_T$ symbols.
\par As illustrated in Fig. 2 and Fig. 3, we assume that at most one URLLC packet per cell and per mini-slot may be generated at an URLLC user in the UL or at the EN in the DL. Each packet is generated at, or intended for, a different URLLC user. The probability of generation of such packet is $a_U$ and packet generation is independent across mini-slots and cells. We note that, in practice, this condition requires an access control protocol for the spectral resources under study that ensures that no more than one URLLC packet is  generated within a cell in a mini-slot.
\par \textit{Channel and Channel State Information (CSI) model:} For both UL and DL, we consider Rayleigh fading channels that are constant over time in the considered frame, but vary independently from one frequency channel to another. Accordingly, the complex channel gain $h^{f}_{i,j}$ between the $i$-th EN and an eMBB user in the $j$-th cell at RU $(f,t)$ is modeled as $h_{i,j}^{f} \sim \cal{CN}$$(0, \alpha^{2}_{i,j})$, where $\alpha^{2}_{i,j}$ accounts for path loss and large scale fading. The channel gains $h_{i,j}^{f}$ are i.i.d. over the frequency index $f$; independent for different pairs $(i,j)$; and constant during the $n_T$ mini-slots of the considered frame. In a similar manner, the channel gains between an URLLC user in cell $i$ and the $i$-th EN are modeled as $g_{i}^{f}(t) \sim \cal{CN}$$(0,\beta_{i}^{2})$, where $\beta_i^2$ reflects the path loss and large-scale fading. Note that the dependence on mini-slot index $t$ is kept for URLLC transmissions in order to highlight that, under the given assumptions, different URLLC users transmit, or are served in each mini-slot $t$. 
\par Following a standard path-loss model, we write $\alpha^{2}_{i,j}= c_B (d_{B,R}/d_{i,j})^{\gamma}$ and $\beta_{i}^{2}=c_U (d_{U,R}/d_{i})^{\gamma}$, where $d_{i,j}$ is the distance between $i$-th EN and the $j$-th eMBB user and $d_i$ is the distance between the $i$-th URLLC user and the EN in the $i$-th cell; $\gamma$ is the path loss exponent and constants $c_B$ and $c_U$ are used to set the signal-to-noise ratio ($\mathrm{SNR}$) levels at the reference distances $d_{B,R}$ and $d_{U,R}$.
% For convenience, we assume that each codeword spans $l$ coherence intervals, i.e., $n_F=L_Fl$. 
\par Due to latency constraints, CSI is assumed to be unavailable at the transmitter side in the communications between a URLLC user and an EN, while receiver CSI is available. This assumption reflects the fact that, in an FDD system, transmitter CSI would require feedback from the receiver. This would limit reliability since it would add another potential cause of error, and it would increase latency. In contrast, CSI is conventionally assumed to be available at both the transmitter and the receiver for the eMBB traffic.
%%%%%%%%%%%%%%%%%%%%%%%%%%%%%%%%%%%%%%%%%%%%%%%%%%%%%%%%%%%%%%%%%%%%%%%%%%%%%%%%%%%%%%%%%%%%%%
\subsection{Heterogeneous Orthogonal and Non-Orthogonal Multiple Access}
In this work, we consider two access schemes, namely H-OMA and H-NOMA. As discussed in Sec. \hyperref[sec:introduction]{I}, these schemes allow the sharing of time and frequency resources between eMBB and URLLC services. As seen in Fig. 2(a), under H-OMA, URLLC packets can only occupy preallocated mini-slots over which eMBB transmissions are not allowed. In particular, a mini-slot is allocated for transmission of URLLC traffic every $L_U$ mini-slots. Parameter $L_U$ hence represents the access latency, i.e., the maximum number of mini-slots a URLLC packet has to wait before transmission. We note that, for the DL, it would also be possible to consider a dynamic schedule of eMBB and URLLC transmissions (see, e.g., \cite{Anand2017JointSO}).
\par In the UL, as illustrated in Fig. 3(a), if multiple URLLC users in a cell generate a packet within the $L_U$ mini-slots between two allocated mini-slots, then a \textit{collision} occurs in the allocated mini-slot. In this case, all packets are discarded due to latency constraints. In the DL, instead, as illustrated in Fig. 3(b), when multiple URLLC packets are generated at an EN, the EN can select one such packet uniformly at random and discard, hence \textit{blocking} from access, all the other packets. Collisions in the UL and blockages in the DL contribute to the overall error rate for URLLC.
\begin{figure}[t]
	\centering
	\includegraphics[height= 4  cm, width= 8 cm]{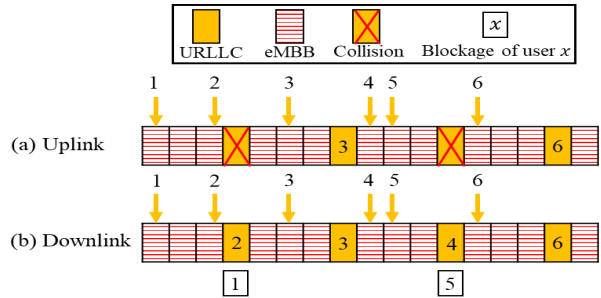}
	\caption{Illustration of (a) collisions in the UL and (b) blockages in the DL for URLLC transmissions under H-OMA when $L_U = 4$. Downward arrows denote the generation of a URLLC packet.}
    \vspace{-5 mm}
	\label{fig:collision_blockage}
\end{figure}
\par In contrast, H-NOMA enables eMBB and URLLC traffic to share the same radio resources. More precisely, as shown in Fig.~\ref{fig:time_frequency}(b), URLLC packets are transmitted in the same mini-slot in which they get generated. It follows that H-NOMA has the minimal access latency of $L_U=1$ at the price of possible interference between eMBB and URLLC signals. Furthermore, under the assumed model, no collisions or blockages occur with H-NOMA.
%%%%%%%%%%%%%%%%%%%%%%%%%%%%%%%%%%%%%%%%%%%%%%%%%%%%%%%%%%%%%%%%%%%%%%%%%%%%%%%%%%%%%%%
\section{Signal Model and Performance Metrics}
\label{sec:signal_model}
In this section, we detail the signal models for UL and DL, as well as the performance metrics of interest. Throughout the analysis, we focus our attention on the analysis of eMBB and URLLC traffic that occupies the frame shown in Fig. 2. We first concentrate on detailing the signal transmitted and received in any of the symbols of an RU $(f,t)$ and then describe the performance metrics of interest. Throughout, we avoid introducing an explicit notation for the indices pointing to one of the $l_F l_T$ symbols in each RU, and hence refer generically with the index $(f,t)$ to any symbol within RU $(f,t)$.
%%%%%%%%%%%%%%%%%%%%%%%%%%%%%%%%%%%%%%%%%%%%%%%%%%%%%%%%%%%%%%%%%%%%%%%%%%%%%%%%%%%%%%%
\subsection{Uplink Signal Model} 
As discussed in the previous section, with H-OMA, one mini-slot is exclusively allocated to URLLC users in the UL every $L_U$ mini-slots (see Fig. 2(a)). We denote as $y_k^{f}(t)$ the signal received by each $\mathrm{EN}_k$ at any symbol within the RU $(f,t)$. This can be written as 
\begin{equation}
y_{k}^{f}(t)=h_{k,k}^{f} x_k^{f}(t) + \sum_{j \neq k}h_{k,j}^{f} x_j^{f}(t) + z_k^{f}(t),
 \label{eq:UL_signalmodel_oma_a}
\end{equation} 
for all $t$ different from $L_U, 2L_U, \ldots$ and where $x_{k}^{f}(t)$ denotes any symbol transmitted in RU $(f,t)$ by the eMBB user that is active in cell $k$ over the given frequency channel $f$; $z_{k}^{f}(t) \sim \mathcal{CN}(0,1)$ is complex Gaussian noise with zero mean and unit variance, which is i.i.d. across the indices $k,f$, and $t$, and across the symbols in an RU. Furthermore, for all mini-slots allocated to URLLC users, the received signal for mini-slot $t=L_U,2L_U ,\ldots$ when there is no collision is
\begin{equation}
y_{k}^{f}(t)=  g_{k}^{f}(t) u_k^f (t) + z_{k}^{f}(t), \label{eq:UL_signalmodel_oma_b}
\end{equation}
where $u_k^{f}(t)$ denotes the signal sent by an URLLC user in the $k$-th cell.
We recall that, in contrast to the eMBB channel, the URLLC channel coefficients $g_{k}^{f}(t)$ depend on the time index $t$ due to the assumption that URLLC packets are generated at different URLLC users.
\par
Unlike H-OMA, under H-NOMA, URLLC users transmit immediately in the mini-slot in which a packet is generated. Accordingly, the signal $y_{k}^{f}(t)$ received at each $\mathrm{EN}_k$ at RU $(f,t)$ can be written as
\begin{equation}
y_{k}^{f}(t)= h_{k,k}^{f} x_k^{f}(t) + \sum_{j\neq k}h_{k,j}^{f} x_{j}^{f}(t) + A_k(t) g_{k}^{f}(t)  u_k^{f}(t) + z_k^{f}(t), \label{eq:UL_signalmodel_noma}
\end{equation}
where $A_k(t) \sim \mathcal{B} \mathbf{\textit{ern}}(a_U)$ is the indicator variable that equals to one if an URLLC packet is generated in cell $k$ at mini-slot $t=1,2,\ldots$.\par
The power constraint for eMBB and URLLC users are defined respectively as
\begin{equation}
\mathbb{E}[|x_{k}^{f}(t)|^{2} ]\leq P_B \label{eq:UL_constraint}\ 
\mathrm{and}\ \mathbb{E}[|u_{k}^{f}(t)|^{2}] \leq P_U,
\end{equation}
where the average in \eqref{eq:UL_constraint} is taken over all uniformly selected information messages.
\par
It will be convenient to write the signal models in matrix form. To this end, the $M \times M$ channel matrix for eMBB users at RU $(f,t)$ is denoted by $\mathbf{H}^f$ with $k$-th row given by the $1 \times M$ vector $ \mathbf{h}_{(k)}^{f}= [h_{k,1}^f,\ldots,h_{k,M}^f]$. The $M \times M$ channel matrix for URLLC users is diagonal due to the discussed lack of inter-cell interference and is denoted as $\mathbf{G}^{f}(t)=\mathrm{diag}(g_{1}^{f}(t), \ldots,g_{M}^{f}(t))$. Consequently, we can write the signals \eqref{eq:UL_signalmodel_noma} received at RU $(f,t)$ across all ENs under H-NOMA as
\begin{equation}
\mathbf{y}^{f}(t) = \mathbf{H}^f \mathbf{x}^{f}(t)+\mathbf{A}(t) \mathbf{G}^{f}(t) \mathbf{u}^{f}(t) + \mathbf{z}^{f}(t), \label{eq:UL_signalmodel_noma_matrix}
\end{equation}
where $\mathbf{A}(t)=\mathrm{diag}(A_1(t),\ldots,A_M(t))$, $\mathbf{x}^{f}(t)= [x_{1}^{f}(t),\ldots ,x_{M}^{f}(t)]^{\mathsf{T}}$, $\mathbf{u}^{f}(t) = [u_{1}^{f}(t), \ldots , u_{M}^{f}(t)]^{\mathsf{T}}$ and $\mathbf{z}^{f}(t) = [z_{1}^{f}(t), \ldots , z_{M}^{f}(t)]^{\mathsf{T}} $. Models \eqref{eq:UL_signalmodel_oma_a}-\eqref{eq:UL_signalmodel_oma_b} can be written in matrix form in an analogous way.
\par Following the general assumptions introduced in Sec. \hyperref[sec:system_model]{II}, the BBU and the ENs are assumed to have available the channel matrices $\mathbf{H}^{f}$ and $\mathbf{G}^{f}(t)$ for both eMBB and URLLC users. Note that providing CSI to the BBU causes a fronthaul overhead that can be considered negligible as the coherence interval size $l_F \times n_T$ increases (see, e.g., \cite{kang_compressionandprecoding_CRAN}). eMBB users are informed by the BBU about the transmission rate at which to operate, while URLLC users have no CSI. As a result, URLLC transmitters adapt their rates only to the distribution of the channel, while the eMBB transmitters adjust their rates to the current channel realization.
%%%%%%%%%%%%%%%%%%%%%%%%%%%%%%%%%%%%%%%%%%%%%%%%%%%%%%%%%%%%%%%%%%%%%%%%%%%%%%%%%%%%%%%%%%%%%%%%%%
\subsection{Downlink Signal Model}
In the DL, for both H-OMA and H-NOMA, the signal $y_k^{f}(t)$ received at an eMBB user in cell $k$ at RU $(f,t)$ can be written as 
\begin{equation}
y_{k}^{f}(t)=  h_{k,k}^f x_k^{f}(t) + \sum_{j \neq k} h_{k,j}^f  x_{j}^{f}(t) + z_k^{f}(t), \label{eq:DL_signalmodel_B}
\end{equation}
where $x_k^{f}(t)$ denotes the symbol transmitted by the $k$-th EN; and $z_k^{f}(t)$ is Gaussian noise received at the eMBB users, with $z_{k}^{f}(t) \sim \mathcal{CN}(0,1)$, which is i.i.d. across the indices $k,f$ and $t$ and across all symbols in an RU. As we will detail in Sec. \hyperref[sec:DL_OMA]{V}, under H-OMA, the signal $x_{k}^{f}(t)$ is either intended for an URLLC user or an eMBB, while for H-NOMA the signal $x_{k}^{f}(t)$ may carry the superposition of URLLC and eMBB signals. We also write \eqref{eq:DL_signalmodel_B} in vector form as
\begin{equation}
y_{k}^{f}(t) = \mathbf{h}_{(k)}^{f} \mathbf{x}^{f}(t) + z_{k}^{f}(t), \label{eq:DL_signalmodel_urllc}
\end{equation}
where $\mathbf{x}^{f}(t)= [x_{1}^{f}(t), \ldots, x_{M}^{f}(t)]^{\mathsf{T}}$ and $\mathbf{h}_{(k)}^{f}=[h_{k,1}^{f},\ldots,h_{k,M}^{f}]$.
\par In contrast, the signal received by the $k$-th URLLC user at RU $(f,t)$ is given as 
\begin{equation}
u_{k}^{f}(t)= g_{k}^{f} (t)  x_{k}^{f}(t) + z_k^{f}(t),\label{eq:W}
\end{equation}
where $z_{k}^{f}(t) \sim \cal{CN}$$(0,1)$ represents Gaussian noise. We recall that, for the same reason as in the UL, the URLLC users' CSI depend on the mini-slot index $t$.
In all cases, the power constraint $P$ for each EN in the DL is defined as
\begin{equation}
\mathbb{E}[|x_{k}^{f}(t)|^{2}] \leq P. \label{eq:dl_constraint_power}
\end{equation}
\par Finally, following the general channels assumptions described in Sec. \hyperref[sec:system_model]{II}, all eMBB and URLLC users are assumed to have available the local channels and signal-to-interference-plus-noise ratio ($\mathrm{SINR}$). The BBU is informed about the eMBB channel matrices $\mathbf{H}^{f}$. Finally, as in the UL scenario the URLLC rate is adjusted to the statistics of the channels, while the eMBB rate is adjusted to the current channel realization.
\par In the remainder of the paper, we will drop the dependence on $t$ when no confusion may arise.
\subsection{Performance Metrics}
\label{sec:signal_model_C}
We are interested in the following performance metrics. For eMBB users, we study the average per-cell sum-rate $R_B$, where the average is taken over the fading distribution. The transmission rates are adapted to the fading realizations thanks to the availability of transmitter CSI. The per-cell sum-rate measures the sum-rate, or, sum-spectral efficiency, in bit/s/Hz across all eMBB users in the system normalized by the number $M$ of cells.
\par As for URLLC users, we define the access latency $L_U$ as the maximum number of mini-slots an URLLC user has to wait before receiving a generated packet. By construction, for H-NOMA, we have $L_U=1$ which corresponds to the minimum access latency when a packet is transmitted in the same mini-slot in which it is generated. Furthermore, following 3GPP\cite[Sec. 7.9]{3gpp_latency}, we define URLLC reliability as the probability to successfully transmit a packet within a given time constraint, here $L_U$.
Accordingly, we explicitly define a constraint on the URLLC error probability $\mathrm{Pr}[\mathcal{E}_U]$ as 
\begin{equation}
\mathrm{Pr}[\mathcal{E}_U] \leq \epsilon_U \label{eq:reliability_constraint}
\end{equation}
for some desired error level $\epsilon_U$. This probability can be interpreted as the average fraction of URLLC devices whose quality-of-service requirements are met.
\par The error event $\mathcal{E}_U$ accounts for two possible types of error, namely collision or blockage and decoding failure. As illustrated in Fig. 3, a collision or a blockage, which only applies to H-OMA, happens in UL or DL respectively, when two or more packets are generated in the $L_U$ mini-slots between two transmission opportunities. In contrast, decoding failure occurs when an URLLC packet is transmitted, and hence it is not subject to collision or blockage, but decoding fails at the receiver. For a given outage probability, due to the absence of CSI at the transmitter, open-loop transmission with no rate adaptation is assumed, and we adopt the maximum transmission rate under an outage probability constraint, or outage capacity, as the rate metric of interest \cite{shortpacketoverrayleighfading}. 
% As proved in \cite{shortpacketoverrayleighfading}, the outage capacity is a suitable measure of the maximum achievable rate for short packet communications as long as the size of the coherence interval $n_T  \times l_F$ is large enough, e.g., larger than around $84$ symbols for $2$ time-frequency diversity branches \cite[Fig. 3]{shortpacketoverrayleighfading}. For smaller coherence intervals, upper and lower bounds on the URLLC rate can be computed as in  \cite{shortpacketoverrayleighfading}. In a manner consistent with eMBB, we define the per-cell sum-outage capacity by summing over all URLLC users. 
\begin{figure*}[t]
	\centering
	\includegraphics[height= 4.5  cm, width= 17cm]{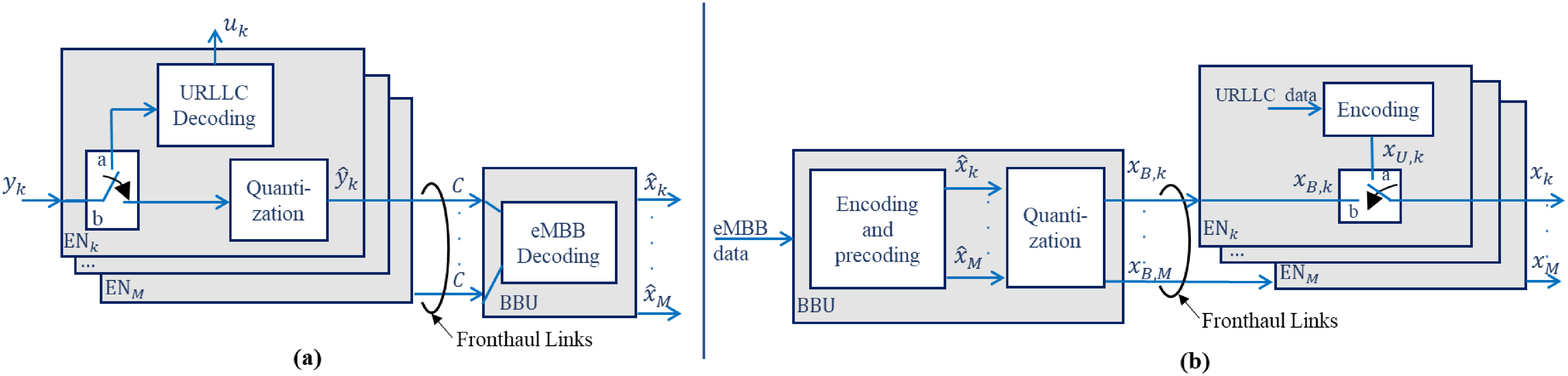}
	\caption{Block diagrams for the ENs and the BBU under H-OMA for (a) UL and (b) DL. Switches in both cases move to position $a$ every $L_U - 1$ mini-slots.}
    \vspace{-5 mm}
	\label{fig:system_model_oma}
\end{figure*}
%%%%%%%%%%%%%%%%%%%%%%%%%%%%%%%%%%%%%%%%%%%%%%%%%%%%%%%%%%%%%%%%%%%%%%%%%%%%%%%%%%%%%%%%%%%%%%%%%
\section{Uplink Orthogonal Multiple Access}
\label{sec:UL_OMA}
In this section, we consider the UL system performance in terms of eMBB rate $R_B$ and URLLC rate $R_U$ for a fixed URLLC access latency $L_U$ and URLLC probability of error requirement $\epsilon_U$ when assuming the conventional H-OMA. The operation of the ENs and of the BBU are illustrated in Fig. 4(a).
%%%%%%%%%%%%%%%%%%%%%%%%%%%%%%%%%%%%%%%%%%%%%%%%%%%%%%%%%%%%%%%%%%
\subsection{URLLC Performance}
As discussed in Sec. II, due to latency constraints, URLLC packets are decoded at the local EN upon reception in the transmission mini-slot $t=L_U,2L_U,...$ (see Fig. \ref{fig:system_model}). For a given decoding error probability $\epsilon_U^D$, the outage capacity of the $k$-th URLLC user is given as
\begin{equation}
R_{U,k}(\epsilon_U^D)= \mathrm{sup} \Big\{R_U: P_{out,k}(R_U) \leq \epsilon_U^D \Big\}, \label{eq:ul_oma_Ru}
\end{equation}
where $P_{out,k}$ denotes the outage probability
\begin{equation}
\begin{aligned}
P_{out,k}(R_U)& = \mathrm{Pr} \Bigg[ \frac{1}{n_F} \sum_{f=1}^{n_F} \mathrm{log}( 1 + S_{U,k}^f) \leq R_U \Bigg], \label{eq:ul_oma_outageproba}
\end{aligned}
\end{equation}
and $S_{U,k}^f =  |g_k^f|^2 P_U$ is the signal-to-noise ratio (SNR) at the EN. The per-cell sum-outage capacity is obtained as $R_U = 1/M \sum_{k=1}^{M} R_{U,k} (\epsilon_{U}^{D})$.
\par Following Sec. \hyperref[sec:signal_model_C]{III.C}, the probability of error of an URLLC packet can be written as
\begin{equation}
\mathrm{Pr}[\mathcal{E}_U] = \sum_{n=1}^{L_U-1} p(n)  + \epsilon_{U}^{D} p(0), \label{eq:error_oma_ul}
\end{equation}
where $p(n)=\mathrm{Pr}[N_U(L_U)=n]$ is the distribution of random variable $N_{U}(L_U) \sim \mathrm{Bin}(L_U-1,a_U)$. The binomial random variable $N_U(L_U)$ represents the number of additional URLLC packets generated by the URLLC users during the remaining $L_U - 1$ mini-slots between two transmission opportunities. The first term in \eqref{eq:error_oma_ul} is the probability that a packet is lost due to collisions, which occurs if $n \geq 1$ additional packets are generated. The second term in \eqref{eq:error_oma_ul} is the probability of a decoding error at the receiver.
%%%%%%%%%%%%%%%%%%%%%%%%%%%%%%%%%%%%%%%%%%%%%%%%%%%%%%%%%%%%%%%%%%%%%%%%%
\subsection{eMBB Rate}
Unlike delay-constrained URLLC traffic, eMBB messages are decoded jointly at the cloud in order to leverage the centralized interference management capabilities of the BBU. To this end, following the standard C-RAN operation, each EN quantizes and compresses the received signal $y_k^f$ for the mini-slots $t \not\in \{L_U,2L_U, ...\}$ by using point-to-point compression (see \cite{zhou2014optimized} \cite{park_robust}\cite{parkandsimeone}), and forwards the resulting signal to the cloud over the fronthaul links (see Fig. 4(a)).
Using \eqref{eq:UL_signalmodel_oma_a}, for each frequency channel $f$, the quantized signal received at the BBU from $\mathrm{EN}_k$ can be written as
\begin{equation}
\hat{y}_{k}^{f}=y_{k}^{f}+q_k^{f},
\end{equation}
where $q_k^{f} \sim \mathcal{CN}(0,\sigma^{2}_{q,k})$ represents the quantization noise with power $\sigma^2_{q,k}$.
From classical results in rate-distortion theory, we have the following relationship between the quantization noise power $\sigma_{q,k}^{2}$ and the fronthaul capacity\cite{el2011network}\cite{parkandsimeone}:
\vspace{-4 mm}
\begin{subequations}
\begin{alignat}{1}
C&=(1- L_U^{-1})\frac{1}{n_{F}} \sum_{f =1}^{n_F} I_{ \mathbf{H}^f}(y_k^f; \hat{y}_k^f)\label{eq:ul_oma_quantizationnoise_a} \\
&=(1-L_U^{-1}) \frac{1}{n_{F}}  \sum_{f =1}^{n_F} \mathrm{log}\Bigg( 1 + \frac{ 1 +  \sum_{j=1}^{M} |h_{k,j}^f|^2 P_B}{\sigma_{q,k}^{2}} \Bigg) .\label{eq:ul_oma_quantizationnoise_b}
\end{alignat} \label{eq:ul_oma_quantizationnoise}%
\end{subequations}
In \eqref{eq:ul_oma_quantizationnoise}, the factor $(1- L_U^{-1})$ captures the fact that a fraction $(1-L_U^{-1})$ of all mini-slots of the wireless channel are occupied by eMBB transmissions. The value of $\sigma^2_{q,k}$ can be obtained by solving \eqref{eq:ul_oma_quantizationnoise_b} via numerical methods.
\par Considering the signals $ \hat{\mathbf{y}}^{f}=[\hat{y}_{1}^{f}, \ldots, \hat{y}_{M}^{f}]$ received by the BBU from all $M$ ENs, the eMBB per-cell sum-rate for a given channel realizations $\mathbf{H}^f$ can be finally written as
\begin{subequations}
\begin{alignat}{1}
R_B&= \frac{(1-L_{U}^{-1})}{M} \frac{1}{n_{F}} \sum_{f=1}^{n_F} I_{ \mathbf{H}^f}(\mathbf{x}^{f};\mathbf{\hat{y}}^{f}) \label{eq:ul_oma_rate_a}\\
&=\! \frac{(1-L_U^{-1})}{M} \frac{1}{n_{F}}\!\! \sum_{f=1}^{n_F} \log \Bigg| \mathbf{I} \!+\! P_B (\mathbf{I}+\mathbf{D}_q)^{-1} \mathbf{H}^f (\mathbf{H}^{f})^{\mathsf{H}} \Bigg| \label{eq:ul_oma_rate_b}
\end{alignat} \label{eq:ul_oma_rate}%
\end{subequations}
where $\mathbf{D}_q=\mathrm{diag}(\sigma_{q,1}^{2}, \ldots,\sigma_{q,M}^{2} )$. The average per-cell sum-rate is obtained by averaging \eqref{eq:ul_oma_rate_b} over the channel realizations $\mathbf{H}^{f}$. A closed form expression for \eqref{eq:ul_oma_rate_b} can be obtained for the case of no fading under the Wyner model \cite{simeone2012cooperative}.
%%%%%%%%%%%%%%%%%%%%%%%%%%%%%%%%%%%%%%%%%%%%%%%%%%%%%%%%%%%%%%%%%%%%%%%%%%%%%%%%%%%%%%%%%%%%
\section{Downlink Orthogonal Multiple Access}
\label{sec:DL_OMA}
In this section, we consider the performance under H-OMA for the DL. Recalling that, as seen in Fig.~\ref{fig:time_frequency}(a), one every $L_U$ mini-slots is allocated to URLLC users, the signal sent by the $k$-th EN for any symbol of the RU $(f,t)$ can be written as
\begin{equation}
x_{k}^{f}(t)=\begin{cases}
x_{U,k}^{f}(t)\ \mathrm{ for }\ t=L_U,2L_U,\ldots\\
x_{B,k}^{f}(t)\ \mathrm{ otherwise},\\
\end{cases}
\end{equation}
where $x_{U,k}^{f}(t)$ and $x_{B,k}^{f}(t)$ are the signals intended for URLLC and eMBB users, respectively, over the given RU. Note that we have $x_{U,k}^{f}(t)=0$ if no URLLC packet was generated in mini-slots $t, t-1, \ldots, t - L_U + 1$.
As a result of the power constraint \eqref{eq:dl_constraint_power}, we have the conditions $\mathbb{E}[|x_{U,k}^{f}(t)|^2] \leq P$ and $\mathbb{E}[|x_{B,k}^{f}(t)|^2] \leq P$.
%%%%%%%%%%%%%%%%%%%%%%%%%%%%%%%%%%%%%%%%%%%%%%%%%%%%%%%%%%%%%%%%%%%%%%%%%%%%%%%%%%%%%%%%%%%%%%
\subsection{URLLC Performance}
The rate analysis of the performance of URLLC traffic under H-OMA in the DL yields the same results \eqref{eq:ul_oma_Ru}-\eqref{eq:ul_oma_outageproba} as for the UL with the caveat that $P_U$ should be replaced by the EN power constraint $P$. This is because in both cases, under H-OMA, URLLC links are interference free. Furthermore, the probability of error \eqref{eq:error_oma_ul} should be modified as 
\begin{equation}
\mathrm{Pr}[\mathcal{E}_U] = \sum_{n=1}^{L_U-1} p(n) \frac{n}{n+1}  +\sum_{n=0}^{L_U-1} p(n) \frac{1}{n+1} \epsilon_{U}^{D}, \label{eq:error_oma_dl}
\end{equation} 
since, in the DL, in case multiple URLLC packets are generated at an EN in the $L_U$ mini-slots per transmission opportunity, one packet can be selected at random and delivered to the corresponding user by the EN. In \eqref{eq:error_oma_dl}, the first term is the probability that more than one additional packets are generated and the packet of interest is blocked from access (see Fig. \ref{fig:collision_blockage}). The second term accounts instead for the decoding error of the transmitted packet.
%%%%%%%%%%%%%%%%%%%%%%%%%%%%%%%%%%%%%%%%%%%%%%%%%%%%%%%%%%%%%%%%%%%%%%%%%%%%%%%%%%%%%%%%%%%%%%%%%
\subsection{eMBB Rate}
Conventional C-RAN transmission based on linear precoding at the BBU and fronthaul quantization is carried out to communicate with eMBB users (see, e.g., \cite{park2013joint}\cite{zhou2016fronthaul}). To elaborate, we define as $s^{f}_{k} \sim \mathcal{CN}(0,1)$ the independent encoded symbols intended for the eMBB user active in cell $k$ over frequency channel $f$ at a given mini-slot. The assumption reflects the use of standard Gaussian random codebooks. As illustrated in Fig. 4(b) the BBU carries out linear precoding separately for each frequency channel $f$, producing the $M \times 1$ vector 
\begin{equation}
\hat{\mathbf{x}}_{B}^{f}=\mathbf{V}^f\mathbf{s}^f, \label{eq:precoding}
\end{equation}
where we have defined the vectors $\mathbf{s}^f=[s^{f}_{1}, s^{f}_{2}, \ldots, s^{f}_{M}]^{\mathsf{T}}$, $\hat{\mathbf{x}}_{B}^{f}=[\hat{x}^{f}_{B,1},\hat{x}^{f}_{B,2},\ldots, \hat{x}^{f}_{B,M}]^{\mathsf{T}}$ and $\mathbf{V}^f$ is the $M \times M$ channel precoding matrix for all eMBB users active on frequency channel $f$. We write $\mathbf{V}^f=[\mathbf{v}_1^f, \ldots, \mathbf{v}_M^f] = [\mathbf{v}_{(1)}^f, \ldots, \mathbf{v}_{(M)}^f]^{\mathsf{T}}$, where $\mathbf{v}_{j}^f \in \mathbb{C}^{M \times 1}$ and $\mathbf{v}_{(j)}^f \in \mathbb{C}^{1 \times M}$ are the $j$-th column and $j$-th row of matrix $\mathbf{V}^f$, respectively.
\par Assuming the standard C-RAN operation where the BBU compresses and forward the eMBB signal to each EN, the signal received at all ENs from the BBU over each frequency channel $f$ can be written as
\begin{equation}
\mathbf{x}_{B}^{f}= \hat{\mathbf{x}}_{B}^{f} + \mathbf{q}^{f}, \label{eq:quantization}
\end{equation}
where we have defined $\mathbf{q}^{f}= [q_{1}^{f}, \ldots,q_{M}^{f}]^{\mathsf{T}}$ and $q_{k}^{f} \sim \mathcal{CN}(0,\sigma^2_{q,k})$ represents the quantization noise with power $\sigma_{q,k}^{2}$. The quantization noise is independent across the EN index $k$ and frequency channel $f$.  Consequently, the received signal \eqref{eq:DL_signalmodel_urllc} at the $k$-th eMBB user can be written as 
\begin{equation}
\begin{aligned}
y^{f}_{k}&= \mathbf{h}^f_{(k)} \sum_{j=1}^{M}  \mathbf{v}_{j}^{f} s^{f}_j + \mathbf{h}^f_{(k)} \mathbf{q}^f + z_{k}^{f}. \label{eq:dl_oma_signalmodel}\\
\end{aligned}
\end{equation}
In order to obtain the quantization noise's power $\sigma^{2}_{q,k}$, in a manner similar to \eqref{eq:ul_oma_quantizationnoise}, we impose the conventional rate distortion condition \cite{el2011network}
\begin{equation}
\begin{aligned}
 C &= (1-L_{U}^{-1}) \frac{1}{n_{F}} \sum_{f = 1}^{n_F} I_{\mathbf{H}^f}(\hat{x}_{B,k}^{f};x^{f}_{B,k}) \\
 & = (1-L_{U}^{-1}) \frac{1}{n_{F}} \sum_{f = 1}^{n_F} \mathrm{log} \Bigg( 1 + \frac{ \|\mathbf{v}_{(k)}^{f} \|^2}{\sigma_{q,k}^{2}}\Bigg)\label{eq:totalC},
\end{aligned}
\end{equation}
for all ENs $k \in {1,\ldots, M} $, which follows from the fronthaul capacity constraint of each EN $k$.
\par
Based on the derivations above, the eMBB achievable per-cell sum-rate for all eMBB users in cell $k$ for given channel realizations $\mathbf{H}^f$ can be written as 
\begin{subequations}
\begin{alignat}{1}
R_{B,k}&=(1-L_{U}^{-1})  \frac{1}{n_{F}} \sum_{f = 1}^{n_F} \log \Bigg(1 + \frac{ |\mathbf{h}^f_{(k)} \mathbf{v}_{k}^f|^2}{1 + \sigma_{\mathrm{eff},k}^{2} }\Bigg). \label{eq:dl_eMBB_H-OMA_b}
\end{alignat}\label{eq:dl_eMBB_H-OMA}%
\end{subequations}
where the effective noise $\sigma_{\mathrm{eff},k}^{2}=\sum_{j=1}^{M}|h_{k,j}^f|^2\sigma^{2}_{q,j} + \sum^{M}_{\substack{j=1\\j \neq k}} |\mathbf{h}^{f}_{(k)} \mathbf{v}_{j}^f|^2$ accounts both for the disturbance due to fronthaul quantization and for eMBB inter-cell interference.\par
Based on the available CSI, precoding can be optimized at the BBU by maximizing the eMBB per-cell sum-rates as
\begin{equation}
\begin{aligned}
& {\text{maximize}}
& &  R_B= \frac{1}{M} \sum_{k=1}^{M} R_{B,k}  \\
& \text{subject to}
& & R_{B,k} \leq \eqref{eq:dl_eMBB_H-OMA_b}\ \ \ \ \ \ \ \ \ \ \ \ \ \ \ \ \ \ \ \ \ \ \ \ \ \  \ \  \ \ \mathrm{for\ all}\ k \\
& & & P \geq \| \mathbf{v}_{(k)}^f \|^2  + \sigma^{2}_{q,k} \ \ \ \ \ \ \ \ \ \ \ \ \ \  \ \mathrm{for\ all}\ k\ \mathrm{and}\ f\\
& & & C \geq  \frac{1-L_U^{-1}}{n_F} \sum_{f=1}^{n_F} \mathrm{log} \Bigg( 1 + \frac{ \|\mathbf{v}_{(k)}^{f} \|^2}{\sigma_{q,k}^{2}}\Bigg) \ \ \mathrm{for\ all}\ k, \label{eq:dl_optim_oma}
\end{aligned}
\end{equation}
where the maximization is over the variables $\{\mathbf{V}^f \}_{f=1}^{n_F}, \{ \sigma_{q,k}^2 \}_{k=1}^{M}, \{R_{B,k}\}_{k=1}^{M}$. The second constraint represents the power constraint at each EN, while the third constraint results from the fronthaul capacity limitations. The problem is non-convex, but it can be tackled using standard methods based on Semidefinite Relaxation (SDR)\cite{relaxation} and Concave-Convex Procedure (CCP) \cite{penaltyccp}. Accordingly, by performing the change of variables $\mathbf{\Omega}_k^f = \mathbf{v}_{k}^{f} (\mathbf{v}_{k}^{f})^{\mathrm{H}}$, adding the constraint $\mathbf{\Omega}_k^f \succeq 0$ and relaxing the constraint $\mathrm{rank}(\mathbf{\Omega}_k^f) =1$, the problem falls in the class of difference of convex problems (DC) \cite{penaltyccp} and thus CCP can be used to solve it as in, e.g., \cite[Sec. IV]{park_joint_optimization}. In order to ensure the rank constraint, we adopt the standard approach of considering the dominant eigen vector $\mathbf{v}_k^f$ of the solution matrices $\mathbf{\Omega}_k^f$ (see, e.g., \cite{relaxation}). 
\begin{figure}[t]
	\centering
	\includegraphics[height= 4.5  cm, width= 8cm]{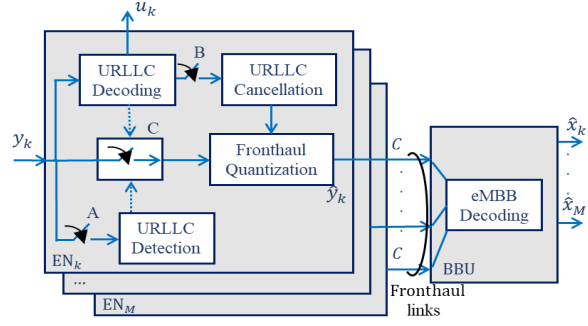}
	\caption{Block diagrams for the ENs and the BBU under H-NOMA for UL with $(i)$ TIN, obtained with switches A and B open and switch C closed; $(ii)$ puncturing, with switch A closed and switch B open; $(iii)$ SIC with switch A open and switch B closed, where switch C remains closed when there is an error in URLLC decoding or detection.}
	\label{fig:system_model_noma_UL}
\end{figure}
\begin{figure}[t]
	\centering
	\includegraphics[height= 3  cm, width= 8.5 cm]{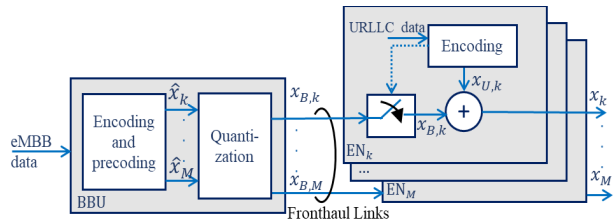}
	\caption{Block diagrams for the ENs and the BBU under under H-NOMA for DL with $(i)$ puncturing, where the switch is open whenever a URLLC packet is encoded and closed otherwise; and $(ii)$ superposition coding, with the switch being always closed.}
    \vspace{-5 mm}
	\label{fig:system_model_noma_DL}
\end{figure}
%%%%%%%%%%%%%%%%%%%%%%%%%%%%%%%%%%%%%%%%%%%%%%%%%%%%%%%%%%%%%%%%%%%%%%%%%%%%%%%%%%%%%%%%%%%%%%%%%%%
\section{Uplink Non-Orthogonal Multiple Access}
\label{sec:UL_NOMA}
In this section we consider the UL performance under H-NOMA. As discussed in Section \hyperref[sec:system_model]{II}, with H-NOMA, eMBB and URLLC users may interfere with each other. For eMBB users, interference is dealt with at the BBU when jointly decoding the eMBB signals. To this end, as illustrated in Fig.~\ref{fig:system_model_noma_UL}, three decoding strategies are studied, namely Treating URLLC Interference as Noise (TIN), puncturing, and Successive Interference Cancellation (SIC). With TIN, ENs quantize and forward the received signals to the BBU on the fronthaul links, and the BBU decodes the eMBB transmissions while treating URLLC signal as noise. Under puncturing, whenever an URLLC user is active in a mini-slot, the receiving EN discards the corresponding eMBB signal received in the same mini-slot prior to quantizing the received signal and forwarding it to the BBU over the fronthaul links. Consequently, the BBU decodes only interference-free eMBB mini-slots. Finally, with the more advanced SIC decoder, the ENs decode and cancel the URLLC transmission before fronthaul quantization. In contrast, for URLLC transmissions, due to reliability and latency constraints, the ENs cannot wait for the entire eMBB frame to be received, and hence URLLC decoding cannot benefit from interference cancellation of the eMBB signal. Therefore, the only affordable decoding strategy for URLLC transmissions is treating eMBB transmissions as noise. 
%%%%%%%%%%%%%%%%%%%%%%%%%%%%%%%%%%%%%%%%%%%%%%%%%%%%%%%%%%%%%%%%%%%%%%%%%%%%%%%%%%%%%%%%%%%%%%%%%
\subsection{URLLC Rate}
With H-NOMA, as illustrated in Fig. 2(b), URLLC users transmit in any mini-slot in which a packet is generated with no additional access latency. We hence have the minimal access latency of $L_U=1$. As for the probability of error, an error can only occur when decoding fails, since no collisions may occur under the given assumptions (see Sec. \hyperref[sec:signal_model]{III}). Hence, the probability of error coincides with the decoding error probability $\epsilon_U^D$ and the reliability constraint \eqref{eq:reliability_constraint} imposes the inequality $\epsilon_{U}^{D} \leq \epsilon_U$. For a given reliability level $\epsilon_U$, the URLLC outage capacity is thus given as in \eqref{eq:ul_oma_Ru} and \eqref{eq:ul_oma_outageproba} with $\epsilon_U^D = \epsilon_U$, and with the caveat that the signal-to-noise-plus-interference ratio $S_{U,k}$ at the $k$-th EN can be written as
\begin{equation}
S_{U,k}^f = \frac{|g_k^f |^2 P_U}{1 + \sum_{j=1}^{M} |h_{k,j}^{f}|^2 P_B}, 
\end{equation}
where $\sum_{j=1}^{M} |h_{k,j}^{f}|^2 P_B$ accounts for interference from eMBB users in all cells active over frequency channel $f$.
%%%%%%%%%%%%%%%%%%%%%%%%%%%%%%%%%%%%%%%%%%%%%%%%%%%%%%%%%%%%%%%%%%%%%%%%%%%%%%%%%%%%%%%%%%%%%%%%%%%%%%%%%%%%%%%%%%
\subsection{eMBB Rate under Treating URLLC Interference as Noise}
Turning to the eMBB performance, we first study the standard C-RAN solution whereby the EN quantizes and forwards all the received signals to the BBU. The BBU decodes the eMBB messages while treating URLLC signals as noise. Under these assumptions, the signal $\mathbf{\hat{y}}^{f} = [\hat{y}_{1}^{f}, \ldots, \hat{y}_{M}^{f}]^{\mathsf{T}}$ received at the cloud from all ENs at RU $(f,t)$ can be written in matrix form using \eqref{eq:UL_signalmodel_noma_matrix}, as 
\begin{equation}
\hat{\mathbf{y}}^f = \mathbf{y}^f  + \mathbf{q}^f =\mathbf{H}^{f} \mathbf{x}^{f} + \mathbf{A} \mathbf{G}^f \mathbf{u}^f + \mathbf{z}^{f}+ \mathbf{q}^{f}, \label{eq:ul_signal_tin}
\end{equation}
where $\mathbf{q}^f = [q_1^f, \ldots, q_M^f]$ and $q_k^f \sim \cal{CN}$$(0, \sigma_{q,k}^{2})$.
In \eqref{eq:ul_signal_tin}, the URLLC activation matrix $\mathbf{A} = \mathrm{diag}(A_1,\ldots ,A_M)$ contains i.i.d. $\mathcal{B}\mathbf{\textit{ern}}(a_U)$ variables. In order to obtain the quantization noises' powers $\sigma_{q,k}^{2}$, in a similar manner as in \eqref{eq:ul_oma_quantizationnoise}, we impose the fronthaul capacity constraint for $k=1,\ldots,M$ as
\begin{equation}
\begin{aligned}
&C=\frac{1}{n_{F}} \sum_{f=1}^{n_F} I_{\mathbf{H}^f , \mathbf{G}^f}( y_{k}^{f}; \hat{y}_{k}^{f}|A_k) \\
&\!\!=\! \frac{1}{n_{F}}\! \sum_{f=1}^{n_F} \mathbb{E}_{A_k}\! \Bigg[ \!\mathrm{log} \Bigg( \!1 \!+ \!\frac{1 + A_k |g_{k}^{f}|^2 P_U + \sum_{j=1}^{M} |h^{f}_{k,j}|^2 P_B }{\sigma_{q,k}^{2}} \!\Bigg) \!\Bigg]\!. \label{eq:ul_TIN_C}
\end{aligned}
\end{equation}
We note that equation \eqref{eq:ul_TIN_C} assumes that the BBU is able to detect the presence of URLLC transmissions. This is reflected in the expectation over the URLLC users' activations.
Finally, the eMBB per-user rate for given channel realizations $\mathbf{H}^f$ is given by \eqref{eq:ul_rate_TIN_b} where we recall the notation $\mathbf{D}_q =\mathrm{diag}(\sigma_{q,1}^{2},\ldots,\sigma_{q,M}^{2})$. The expectation in \eqref{eq:ul_rate_TIN_b} can in practice be computed exactly by summing over the $2^M$ possible values for matrix $\mathbf{A}$ as long as $M$ is not too large. Otherwise, stochastic approximation methods can be used.
\newcounter{storeeqcounter_three}
\newcounter{tempeqcounter3}
\addtocounter{equation}{1}
\setcounter{storeeqcounter_three}
{\value{equation}}
\begin{figure*}[!t]
\normalsize
\setcounter{tempeqcounter3}{\value{equation}} % temp store of current value
\begin{IEEEeqnarray}{rCl}
\setcounter{equation}{\value{storeeqcounter_three}} % number of this equation
R_B&= \frac{1}{M n_{F}} \sum_{f=1}^{n_F} I_{\mathbf{H}^f, \mathbf{G}^f}(\mathbf{x}^{f};\mathbf{\hat{y}}^{f}| \mathbf{A}) = \frac{1}{M n_{F}} \sum_{f=1}^{n_F} \mathbb{E}_{ \mathbf{A}} \Bigg[ \mathrm{log} \frac{\mathrm{det}  \Big( \mathbf{I} + \mathbf{D}_q + P_B \mathbf{H}^f (\mathbf{H}^f)^{\mathsf{H}} +P_U \mathbf{A} \mathbf{G}^f (\mathbf{G}^f)^{\mathsf{H}} \Big)}{\mathrm{det} \Big( \mathbf{I} + \mathbf{D}_q + P_U \mathbf{A} \mathbf{G}^f (\mathbf{G}^f)^{\mathsf{H}}\Big)} \Bigg].
\label{eq:ul_rate_TIN_b}\end{IEEEeqnarray}
\setcounter{equation}{\value{tempeqcounter3}} % restore correct value
\hrulefill
\end{figure*}
%%%%%%%%%%%%%%%%%%%%%%%%%%%%%%%%%%%%%%%%%%%%%%%%%%%%%%%%%%%%%%%%%%%%%%%%%%%%%%%%%%%%%%%%%%%%%%%%%%%%%%%%%%%%%%%
\subsection{eMBB Rate under Puncturing}
\label{sec:UL_NOMA_C}
With puncturing, as seen in Fig.~\ref{fig:system_model_noma_UL}, whenever an URLLC user is active in a given cell $k$, and RU $(f,t)$, the signal $y_{k}^{f}(t)$ received at the EN is discarded and not forwarded to the BBU. Note that, with the assumed grant-free URLLC transmissions, this requires the detection of URLLC user's activity prior to fronthaul quantization, e.g., based on the detection of URLLC references sequences. A similar approach is under consideration within 3GPP \cite{3gpp_meeting92}.\par
To elaborate, we assume that each EN detects correctly that there are transmissions of URLLC devices. The assumption is well justified by the high reliability of URLLC transmissions. The EN compresses and forwards only the signals received during mini-slots free of interference from URLLC transmissions. Under this assumption, the signal $\tilde{y}^{f}_{k}$ received at the cloud from $\mathrm{EN}_k$ over RU $(f,t)$ can be written as
\begin{equation}
\tilde{y}_k^{f}=(1-A_k)\bigg(h_{k,k}^{f} x_k^{f} + \sum_{j\neq k}h_{k,j}^{f} x_{j}^{f} \bigg)+ z_k^{f} + q_k^f. \label{eq:UL_signalmodel_noma_puncturing} 
\end{equation}
According to \eqref{eq:UL_signalmodel_noma_puncturing}, the received signal $\tilde{y}_k^{f}(t)$ carries no information, i.e., only noise, when an URLLC user is active ($A_k=1$) in the corresponding mini-slot. Otherwise, when $A_k=0$, the signal contains the contributions of the eMBB users and of the quantization noise $q_k^f (t) \sim \cal{CN}$$(0,\sigma_{q,k}^{2})$.
In matrix form, the signal in \eqref{eq:UL_signalmodel_noma_puncturing} received across all ENs over RU $(f,t)$ can be equivalently written as
\begin{equation}
\tilde{\mathbf{y}}^f = (\mathbf{I}-\mathbf{A}) \mathbf{H}^{f} \mathbf{x}^{f} + \mathbf{z}^{f}+ \mathbf{q}^{f} \label{eq:UL_signalmodel_noma_puncturing_matrix}.
\end{equation} 
\par In order to enable decoding, the BBU at the cloud needs to be informed not only of the signals \eqref{eq:UL_signalmodel_noma_puncturing} for all the mini-slots with $A_k=0$ for all ENs $k$, but also of the location of such mini-slots. To this end, each $\mathrm{EN}$ collects the i.i.d. binary vector containing the $n_T$ Bernoulli variables $A_k \sim \mathcal{B} \mathbf{\textit{ern}}(1-a_U)$. The number of bits needed to be communicated from $\mathrm{EN}_k$ to the BBU in order to ensure the lossless reconstruction of this sequence is given by $n_T H(a_U) \ \mathrm{bits},$ where $H(a_U) = -a_U\log a_U - (1-a_U)\log(1-a_U)$ is the binary entropy function \cite{el2011network}.
Based on the discussion above, imposing fronthaul capacity constraint yields the condition 
\begin{equation}
\begin{aligned}
n_T n_F  l_T l_F C &= n_T  l_T l_F (1-a_U)\\
&\times \sum_{f =1}^{n_F} I_{\mathbf{H}^f}(y_{k}^{f};\tilde{y}_{k}^{f}| A_k = 0) + n_T H(a_U),
\end{aligned}
\label{eq:ul_noma_puncturing_totalC}
\end{equation}
where $I_{\mathbf{H}^f}(y_{k}^{f};\tilde{y}_{k}^{f}| A_k = 0)=\mathrm{log}(1 + (1 + \sum_{j=1}^{M} |h_{k,j}^f|^2 P_B)/\sigma_{q,k}^2 )$ and we recall that $n_T n_F  l_T l_F C$ is the total number of bits per frame available for transmission on each fronthaul link, and the mutual information term accounts for the compression of the received signals over the $n_T n_F l_T l_F  (1-a_U)$ symbols unaffected by URLLC interference (i.e., with $A_k = 0$).
The quantization noise's powers $\sigma_{q,k}^2$ can be obtained by solving \eqref{eq:ul_noma_puncturing_totalC} using numerical means for all $k=1, \ldots , M$. Following \eqref{eq:ul_rate_TIN_b}, the eMBB per user rate for given channel realization $\mathbf{H}^f$ is finally given by
\begin{equation}
R_B\!= \!\frac{1}{M n_F} \! \sum_{f =1}^{n_F} \mathbb{E}_{\mathbf{A}} \Bigg[ \!\log \Bigg|\mathbf{I} + P_B (\mathbf{I} + \mathbf{D}_q)^{-1} (\mathbf{I}-\mathbf{A}) \mathbf{H}^f \! (\mathbf{H}^{f})^{\mathsf{H}} \Bigg| \Bigg]\!. \label{eq:ul_noma_puncturing_rate_b}%
\vspace{-2 mm}
\end{equation}
A closed form expression for \eqref{eq:ul_oma_rate_b} can be obtained for the case of no fading under the Wyner model \cite{simeone2012cooperative} \cite{rahifuplink}.
%%%%%%%%%%%%%%%%%%%%%%%%%%%%%%%%%%%%%%%%%%%%%%%%%%%%%%%%%%%%%%%%%%%%%%%%%%%%%%%%%%%%%%%%%%%%%%%%%%%%%%%%%%%%%%%%%
\subsection{eMBB Rate under Successive Interference Cancellation of URLLC Users}
\label{sec:DL_NOMA}
We finally study a more complex receiver architecture whereby SIC of URLLC packets is carried out at the ENs prior to fronthaul quantization. More specifically, as seen in Fig.~\ref{fig:system_model_noma_UL}, if an URLLC user is active and its message is decoded correctly at the receiving EN, the URLLC message is canceled by the EN. If decoding is unsuccessful, the signal received in the corresponding mini-slots is treated as in puncturing. Accordingly, with this scheme, each EN quantizes the received signals only for the minislots that are either free of URLLC transmissions or that contain URLLC messages that were successfully decoded and canceled at the EN.
As a result, the received signal at the BBU from $k$-th EN can be written as \eqref{eq:UL_signalmodel_noma_puncturing} but with an erasure probability of $a_U\epsilon_{U}^{D}$ instead of $a_U$. This is because a mini-slot is dropped if an URLLC user in the cell is active, which happens with probability $a_U$, and if its transmission is decoded incorrectly, which happens with probability $\epsilon_U^D$. As a result, the eMBB rate can be evaluated as \eqref{eq:ul_noma_puncturing_rate_b} with the caveat that the random variables $A_k$ for $k=1, \ldots, M$ are i.i.d. $\mathcal{B} \mathbf{\textit{ern}}$$(a_U \epsilon_{U}^{D})$.
%%%%%%%%%%%%%%%%%%%%%%%%%%%%%%%%%%%%%%%%%%%%%%%%%%%%%%%%%%%%%%%%%%%%%%%%%%%%%%%%%%%%%%%%%%%%%%%%%%%%%%%%%%%%%
\section{Downlink Non-Orthogonal Multiple Access}
\label{sec:DL_NOMA}
In this section we consider H-NOMA for the DL scenario. We follow an approach similar to the UL in Sec. \hyperref[sec:DL_OMA]{V} by allowing for different interference management strategies between URLLC and eMBB. A key new aspect in the DL is that interference arises from the URLLC transmissions originating at the ENs, which are a priori unknown to the BBU. Accordingly, as illustrated in Fig.~\ref{fig:system_model_noma_DL}, we consider two interference management strategies at the ENs, namely puncturing and superposition coding. Under puncturing, in any mini-slot in which an URLLC packet is generated, the EN transmits only the URLLC packet, dropping any eMBB information. Instead, with superposition coding, the EN transmits a superposition of both eMBB and URLLC signals to both users.
%%%%%%%%%%%%%%%%%%%%%%%%%%%%%%%%%%%%%%%%%%%%%%%%%%%%%%%%%%%%%%%%%%%%%%%%%%%%%%%%%%%%%%%%%%%%%%%%%%%%%%%%%%%%
\subsection{Puncturing}
For both puncturing and superposition coding, the BBU precodes the eMBB signals and forwards them to the ENs over the fronthaul links. Under puncturing, whenever an URLLC packet is generated at an EN in a given mini-slot, the EN discards the eMBB signal received for the same mini-slot from the BBU. Note that this does not affect the scheduling decision made at the BBU for eMBB traffic, whose packet still spans the frame, with the exclusion of the punctured mini-slots. Consequently, the transmitted signal $x_{k}^{f}$ by $\mathrm{EN}_k$ can be written as
\begin{equation}
x_{k}^{f} = (1-A_k) (x_{B,k}^{f} + q_{k}^{f}) + A_k x_{U,k}^{f},
\end{equation}
where we recall that $A_k  \sim \mathcal{B} \mathbf{\textit{ern}}(a_U)$ is the binary random variable denoting  the generation of a URLLC packet at the EN in mini-slot $t$.
\par \textit{URLLC Rate:} URLLC users' outage capacity is the same as in the H-OMA case discussed in Sec. \hyperref[sec:DL_OMA]{V} due the absence of inter-cell interference at URLLC users. However, with H-NOMA, the probability of error is equal to the decoding error probability due to the absence of collisions between URLLC packets. As a result, the URLLC rate is given by \eqref{eq:ul_oma_Ru} with $\epsilon_{U}^{D} = \epsilon_U$.
% Under this assumption, the signal $\tilde{Y}^{f}_{k}$ received at the $k$-th eMBB user on frequency $f$ can be written as
% \begin{equation}
% \begin{aligned}
% \tilde{Y}_k^{f}=B_k X_k^{f} + \alpha B_{k-1} X_{k-1}^{f} + \alpha B_{k+1} X_{k+1}^{f} + N_k^{f}  ,\label{eq:7}
% \end{aligned}
% \end{equation}
% where the random variable $B_k=1-A_k\sim \mathcal{B}(1-q)$ indicates the absence ($B_k=1$) or presence ($B_k=0$) of a URLLC packet generated in the given minislot by the $k$-th EN. According to \eqref{eq:7}, the received signal $\tilde{Y}_k^{f}$ carries no information for the eMBB user when the same cell URLLC user is active ($B_k=0$).
\par \textit{eMBB Rate: }By assuming linear precoding at the BBU with precoding matrix $\mathbf{V}^f$ as in Sec. \hyperref[sec:DL_OMA]{V}, the signal received by the $k$-th eMBB user in a given mini-slot can be written in a manner similar to \eqref{eq:dl_oma_signalmodel} as
\begin{equation}
y_{k}^{f}= \mathbf{h}^{f}_{(k)} (\mathbf{I}- \mathbf{A}) \Big(\sum_{j=1}^{M} \mathbf{v}_j^f s_{j}^{f}  + \mathbf{q}^{f} \Big) + \mathbf{h}_{(k)}^{f} \mathbf{A}  \mathbf{x}_{U}^{f} + n^{f}_{k}, \label{eq:dl_noma_puncturing_e}
\end{equation}
with definitions given in Sec. \hyperref[sec:signal_model]{III}.
According to \eqref{eq:dl_noma_puncturing_e}, an eMBB user receives useful  information only from the ENs in cells $j$ for which no URLLC packet is generated i.e., $A_j=1$. The per-user eMBB rate at the $k$-th user and for given channel realizations $\mathbf{H}^f$ can be written as \eqref{eq:dl_puncturing_embb}
% \begin{subequations}
% \begin{alignat}{1}
% R_{B,k}&= \frac{1}{n_F} \sum_{f =1}^{n_F} I_{\mathbf{H}^{f}}(s_{k}^{f}; y_{k}^{f}|\mathbf{A}) \label{eq:dl_puncturing_embb_a}\\
% &=\frac{1}{n_F} \sum_{f =1}^{n_F} \mathbb{E}_{\mathbf{A}} \Bigg[\log \Bigg(1 + \frac{|\mathbf{h}^{f}_{(k)} (\mathbf{I}-\mathbf{A}) \mathbf{v}_k^f|^2}{1+ \sum_{j=1}^{M} |h_{k,j}^f|^2 ((1-A_j) \sigma^{2}_{q,j} + A_j P )+ \sum^{M}_{\substack{j=1\\j \neq k}} |\mathbf{h}^{f}_{(k)}  (\mathbf{I}-\mathbf{A}) \mathbf{v}_j^f|^2  }\Bigg) \Bigg],\label{eq:dl_puncturing_embb_b}
% \end{alignat}\label{eq:dl_puncturing_embb}
% \end{subequations}
\newcounter{storeeqcounter_one}
\newcounter{tempeqcounter}
where the term $\sum_{j=1}^{M} |h_{k,j}^f|^2 A_j P$ accounts for URLLC interference and $\sum^{M}_{\substack{j=1\\j \neq k}} |\mathbf{h}^{f}_{(k)}  (\mathbf{I}-\mathbf{A}) \mathbf{v}_j^f|^2$ accounts for eMBB interference. We remark that achievability of \eqref{eq:dl_puncturing_embb} requires the capability of eMBB users to detect URLLC transmissions, e.g., using a specific preamble in the URLLC mini-slots. Furthermore, we note that, with puncturing, eMBB and URLLC transmissions are orthogonal within each cell, but inter-service interference still arises due to the asynchronous URLLC packet generation across cells.
\addtocounter{equation}{1}
\setcounter{storeeqcounter_one}
{\value{equation}}

\begin{figure*}[!t]
\normalsize
\setcounter{tempeqcounter}{\value{equation}} % temp store of current value
\begin{IEEEeqnarray}{rCl}
\setcounter{equation}{\value{storeeqcounter_one}} % number of this equation
R_{B,k}&= &\frac{1}{n_F} \sum_{f =1}^{n_F} \mathbb{E}_{\mathbf{A}} \Bigg[\log \Bigg(1 + \frac{|\mathbf{h}^{f}_{(k)} (\mathbf{I}-\mathbf{A}) \mathbf{v}_k^f|^2}{1+ \sum_{j=1}^{M} |h_{k,j}^f|^2 ((1-A_j) \sigma^{2}_{q,j} + A_j P )+ \sum^{M}_{\substack{j=1\\j \neq k}} |\mathbf{h}^{f}_{(k)}  (\mathbf{I}-\mathbf{A}) \mathbf{v}_j^f|^2  }\Bigg) \Bigg],
\label{eq:dl_puncturing_embb}
\end{IEEEeqnarray}
\setcounter{equation}{\value{tempeqcounter}} % restore correct value
\hrulefill
\vspace*{4pt}
\end{figure*}
\par
In a manner similar to \eqref{eq:dl_eMBB_H-OMA}, optimal linear precoding can be carried out by maximizing the sum-rates at the BBU. An interesting aspect of this problem is that while the channel matrix is known at the BBU, the effective channels $ (\mathbf{I}-\mathbf{A}) \mathbf{H}^f$ are not known to the BBU due to the presence of the random URLLC activation matrix $\mathbf{A}$. The sum-rate maximization problem can be formulated and tackled in a manner similar to \eqref{eq:dl_optim_oma}.
%%%%%%%%%%%%%%%%%%%%%%%%%%%%%%%%%%%%%%%%%%%%%%%%%%%%%%%%%%%%%%%%%%%%%%%%%%%%%%%%%%%%%%%%%%%%%%%
\subsection{Superposition Coding}
Under this strategy, each $\mathrm{EN}$ transmits a superposition of the signal $x_{U,k}^{f}$ intended for URLLC users and the signal $x_{B,k}^{f}$ for eMBB users. We fix the power of the signal intended to the URLLC user to $\mathbb{E}[| x^{f}_{U,k} |^2] = P_U \leq P$. When designing the beamforming matrices $\mathbf{V}^f$, the BBU assumes an available power of $P$ since it is not aware of the URLLC activations $A_k$ for $k=1,\ldots ,M$. We hence have the constraint $\| \mathbf{v}_{(k)}^{f} \|^2 \leq P - \sigma_{q,k}^{2}$ as for puncturing. Accordingly, the signal $x_{k}^{f}$ transmitted by the $k$-th EN over RU $(f,t)$ can be written as 
\begin{equation}
x_{k}^{f}= (1 + A_k (\sqrt{\delta} - 1) )(x_{B,k}^{f} + q_{k}^{f})+ A_k x_{U,k}^{f},\label{eq:x_urllcasnoise}
\end{equation}
with the scaling factor $\delta= 1 - P_U/ P $.
The signal \eqref{eq:x_urllcasnoise} is such that, when $A_k=0$, only the eMBB signal $x_{B,k}^{f} + q_{k}^{f}$ is transmitted; and, when $A_k=1$, the transmitted signal is given by the superposition $\sqrt{\delta} (x_{B,k}^{f} + q_{k}^{f}) + x_{U,k}^{f}$, where the factor $\delta$ guarantees the EN power constraint. Note that this strategy is a generalization of puncturing which is obtained by setting $P_U=P$ in \eqref{eq:x_urllcasnoise}.
\par
\textit{URLLC Performance:} The URLLC rate and corresponding outage probability can be obtained using \eqref{eq:ul_oma_Ru} and \eqref{eq:ul_oma_outageproba} by setting the following value of the signal-to-noise-plus-interference ratio at the $k$-th EN
\begin{equation}
S_{U,k}^f = \frac{| g_{k}^{f}|^2 P_U}{1 + | g_{k}^{f}|^2 (P-P_U)},
\end{equation}
where the disturbance term $|g_{k}^f|^2 (P-P_U)$ represents the eMBB interference towards the URLLC user.
\par \textit{eMBB Performance:} The signal $y_{k}^{f}$ received at the $k$-th eMBB user on frequency $f$ can be written as
\begin{equation}
y_{k}^{f}= \mathbf{h}_{(k)}^f \Bigg( \Big( \mathbf{I} + \mathbf{A}(\mathbf{\Delta} - \mathbf{I})\Big) \Big( \sum_{j=1}^{M} \mathbf{v}_j^f s_{j}^{f}  +\mathbf{q}^{f} \Big) + \mathbf{A} \mathbf{x}_{U}^{f} \Bigg) + z^{f}_{k}, \label{eq:dl_noma_superposition_signal}
\end{equation}
where $\mathbf{\Delta} = \delta \mathbf{I}$.
\newcounter{storeeqcounter_two}
\newcounter{tempeqcounter2}
\addtocounter{equation}{1}
\setcounter{storeeqcounter_two}
{\value{equation}}
\begin{figure*}[!t]
\normalsize
\setcounter{tempeqcounter2}{\value{equation}} % temp store of current value
\begin{IEEEeqnarray}{rCl}
\setcounter{equation}{\value{storeeqcounter_two}} % number of this equation
R_{B,k}&=& \frac{1}{n_F} \sum_{f=1}^{n_F} \mathbb{E}_{\mathbf{A}} \Bigg[ \log \Bigg(1 + \frac{|\mathbf{h}^{f}_{(k)} ( \mathbf{I} + \mathbf{A}(\mathbf{\Delta} - \mathbf{I})) \mathbf{v}_k^f|^2}{1+ \sum_{j=1}^{M} |h_{k,j}^f|^2(W_j \sigma_{q,j}^{2} + A_j P_U)+\sum^{M}_{\substack{j=1\\j \neq k}}  |\mathbf{h}^{f}_{(k)} ( \mathbf{I} + \mathbf{A}(\mathbf{\Delta} - \mathbf{I})) \mathbf{v}_j^f|^2} \Bigg) \Bigg].\label{eq:dl_noma_superposition_e}%
\end{IEEEeqnarray}
\setcounter{equation}{\value{tempeqcounter2}} % restore correct value
\hrulefill
\vspace*{2pt}
\end{figure*}
Assuming again that the eMBB users can detect URLLC packets, the eMBB per-user rate can be written as \eqref{eq:dl_noma_superposition_e} where $W_k =1 + A_k (\delta - 1)$. The sum-rate maximization problem can be formulated and tackled in a manner similar to \eqref{eq:dl_optim_oma}.
\begin{figure}[ht]
	\centering
	\includegraphics[height= 5.5 cm, width= 9 cm]{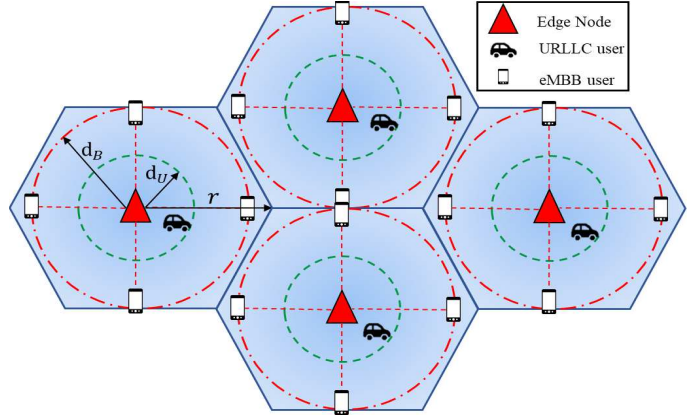}
	\caption{Simulations setup with $M=4$ cells, four eMBB users and one URLLC user per-cell, a cell radius of $r=2\ \mathrm{km}$ and an URLLC zone (small green circles) with radius of $d_U=0.1\ \mathrm{km}$.}
    \vspace{-5 mm}
	\label{fig:system_model_simulation}
\end{figure}
\section{Numerical Results}
\label{sec:numerical_results}
\subsection{Simulations Set-Up}
As illustrated in Fig.~\ref{fig:system_model_simulation}, we consider a system of $M=4$ cells where each cell contains four eMBB users and one URLLC user. The cells' radius is $r=2\ \mathrm{km}$. In order to focus on worst-case performance guarantees, eMBB users are located on the circle of radius $d_B=r \mathrm{sin}(\pi/3)\ \mathrm{km}$, as shown in Fig.~\ref{fig:system_model_simulation}. In contrast, URLLC users are arbitrary placed on the circle of radius $d_U=0.1\ \mathrm{km}$. Furthermore, we set $n_F=4$ frequency channels in total, with $n_F^B=1$ frequency channel for each eMBB user. The constraint on the URLLC probability of error is $\epsilon_U=10^{-3}$. As in 3GPP release 15 \cite{3gpp_release15}, we set $l_F=12$ subcarriers and $l_T=14$ symbols. For the UL, the eMBB power $P_B$ is fixed to $6.4\ \mathrm{dBm}$, while the URLLC power is $P_U=23\ \mathrm{dBm}$. As for the DL, the transmission power of each EN is set to $P=24.77\ \mathrm{dBm}$, and, for superposition coding, we fix $P_U = 23\ \mathrm{dBm}$. The constant $c_{B}$ in the path loss formula is chosen so as to obtain an average SNR of $3\ \mathrm{dB}$ for eMBB at the reference distance $d_{B,R}=d_B$ for both UL and DL with transmission powers $P_B$ in the UL and $P$ in the DL\cite{lte_power}. The constant $c_{U}$ is instead chosen so as to obtain an average URLLC SNR equal to $10\ \mathrm{dB}$ at the reference distance $d_{U,R}=d_U$ for both UL and DL, with transmission powers $P_U=23\ \mathrm{dBm}$ for both UL and DL\cite{lte_power}. Finally, unless stated otherwise, we assume throughout this section the values $C=2\ \mathrm{bit/s/Hz}$, $\gamma=3$, $a_U=0.5\times10^{-3}$ and $L_U=2$ for H-OMA. 
% We note that the maximum transmit power of a user equipment in LTE is $23\ \mathrm{dBm}$\cite{lte_power} and the average transmit power is around $6.4\ \mathrm{dBm}$ for $95\%$ of the time\cite{lte_power}
% \footnote{\textcolor{blue}{In 3GPP \cite[Sec. 7.9]{3gpp_latency}, the probability of error of a packet over the air interface should be around $10^{-5}$, this value is for a frame duration of $10\ \mathrm{ms}$, i.e., $80$ mini-slots. This value is for ultra reliable but not low latency packets \cite[Sec. 7.5]{3gpp_latency} of a duration of 1 mini-slot.}}
%%%%%%%%%%%%%%%%%%%%%%%%%%%%%%%%%%%%%%%%%%%%%%%%%%%%%%%%%%%%%%%%%%%%%%%%%%%%%%%%%%%%%%%%%
\begin{figure}[t]
	\centering
	\includegraphics[height= 6.3 cm, width= 9.5 cm]{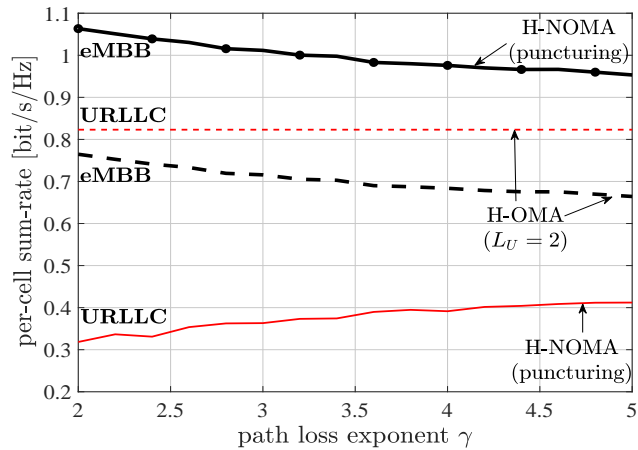}
	\caption{UL eMBB average per-cell sum-rate and URLLC per-cell sum-outage capacity as a function of the path loss exponent $\gamma$ under H-OMA with $L_U=2$ and H-NOMA with puncturing for the setup in Fig.~\ref{fig:system_model_simulation} ($\epsilon_U=10^{-3},a_U=0.5\times 10^{-3}$, $C=2\ \mathrm{bit/s/Hz}$)}
    \vspace{-5 mm}
	\label{fig:UL_function_gamma}
\end{figure}
\subsection{Uplink}
We first consider the UL. In Fig.~\ref{fig:UL_function_gamma}, we plot the eMBB and URLLC per-cell sum-rate as function of the path loss exponent $\gamma$. The average power of the direct channels from each eMBB user to the same cell EN are independent of $\gamma$ due to the assumption that the reference distance $d_{B,R}$ coincides with the distance $d_{B}$. Consequently, decreasing the value of $\gamma$ effectively increases the inter-cell channel gains for eMBB users (see Sec. \hyperref[sec:system_model]{II}). For H-NOMA, we consider the simplest form of processing, namely puncturing, as studied in Sec. \hyperref[sec:UL_NOMA_C]{VI.C}. We first observe that, in the given scenario with small $a_U$, H-OMA offers a higher URLLC transmission rate due to the absence of interference of eMBB users, but this comes at the price of the higher URLLC access latency of $L_U=2$ mini-slots. In contrast, H-NOMA provides the minimal access latency of $L_U=1$, while supporting a lower URLLC rate that decreases for lower values of $\gamma$ due to the increasing eMBB interference power. 
Furthermore, for eMBB traffic, H-NOMA provides a larger rate due to the larger number of available mini-slots. Finally, under both H-NOMA and H-OMA, the eMBB rate increases for decreasing $\gamma$ thanks to the joint decoding carried out at the BBU, which can benefit from the inter-cell signal paths.
\begin{figure}[t]
	\centering
	\includegraphics[height= 6.3 cm, width= 9.5 cm]{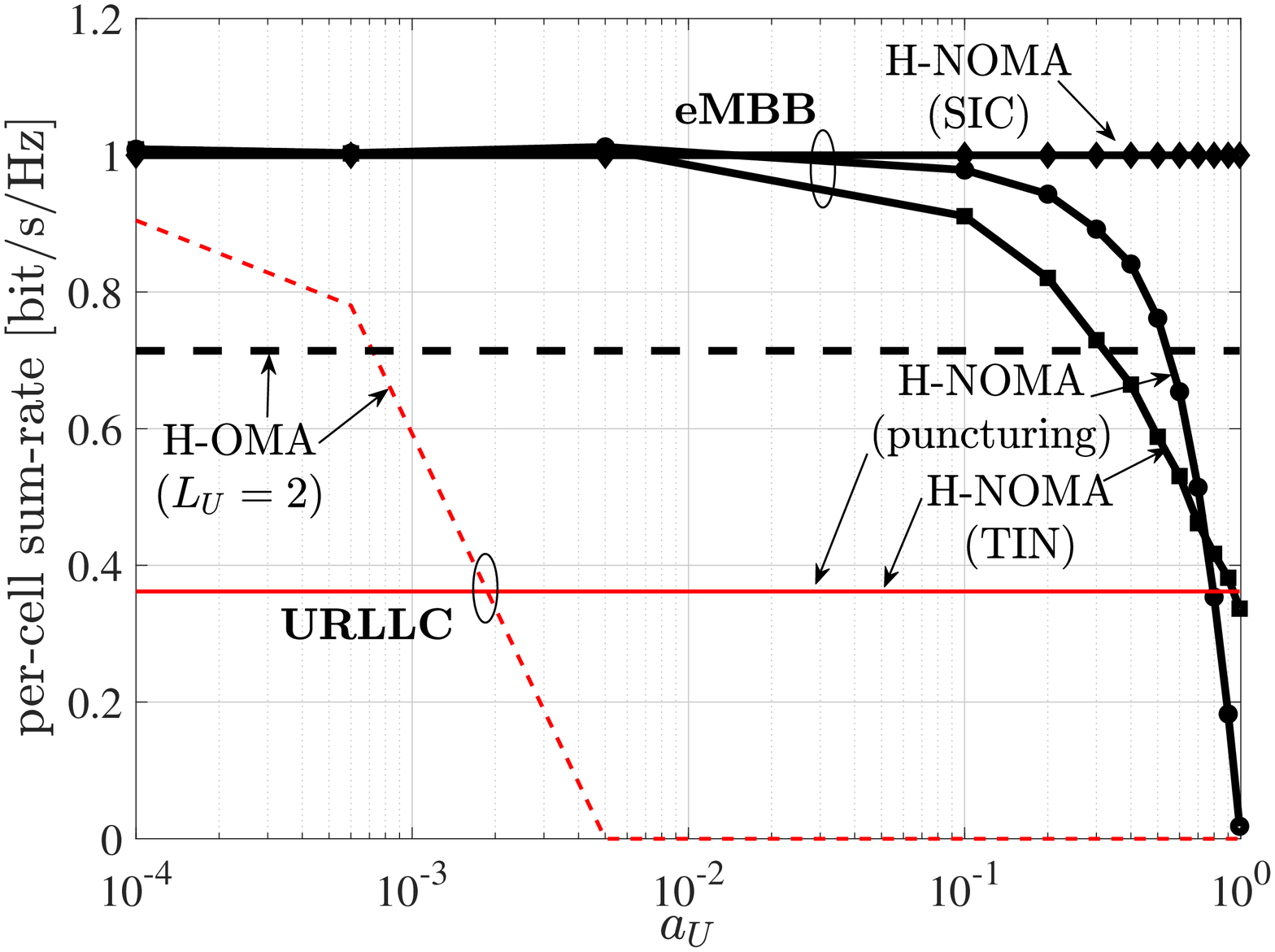}
	\caption{UL eMBB average per-cell sum-rate and URLLC per-cell sum-outage capacity as a function of the packet generation probability $a_U$ for URLLC traffic for H-OMA with $L_U=2$ and for H-NOMA with different decoding strategies for the set-up in Fig.~\ref{fig:system_model_simulation} ($\epsilon_U = 10^{-3},C=2\ \mathrm{bit/s/Hz}$, $\gamma=3$)}
    \vspace{-5 mm}
	\label{fig:UL_function_a_U}
\end{figure}
\par In Fig.~\ref{fig:UL_function_a_U}, we further investigate the per-user eMBB and URLLC rates as a function of URLLC traffic generation probability $a_U$. The URLLC users' rate under H-OMA is seen to decrease quickly as a function of $a_U$. This is because, as $a_U$ increases, the error probability in \eqref{eq:error_oma_ul} becomes limited by the probability that an URLLC packet is undergoes a collision due to an insufficient number of transmission opportunities. For H-NOMA, the URLLC rate is instead not affected by $a_U$. As for eMBB, for $a_U \leq 0.86$, treating URLLC signals as noise achieves the worst eMBB rate among the H-NOMA schemes. In fact, in this regime, if the fronthaul capacity is small, it is preferable not to waste fronthaul resources by quantizing samples affected by URLLC interference. In contrast, for larger values of $a_U$, puncturing becomes the worst-performing H-NOMA strategy, since the achievable eMBB rate becomes limited by the small number of useful received signal samples forwarded to the BBU. Finally, the more complex SIC scheme always provides the largest per-user eMBB rate thanks to the high probability of cancellation of URLLC signals at the EN.\par
\begin{figure}[t]
	\centering
	\includegraphics[height= 6.3 cm, width= 9.5 cm]{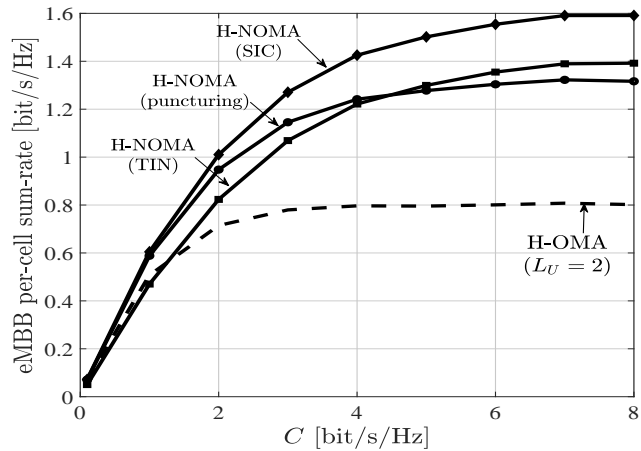}
	\caption{UL eMBB average per-cell sum-rate as a function of the fronthaul capacity $C$ for H-OMA with $L_U=2$ and for H-NOMA with different decoding strategies for the set-up in Fig.~\ref{fig:system_model_simulation} ($\epsilon_U = 10^{-3},\gamma=3$, $a_U=0.2$)}
    \vspace{-5 mm}
	\label{fig:UL_function_C}
\end{figure}
In Fig.~\ref{fig:UL_function_C}, we plot the eMBB per-cell sum-rate rate as a function of the fronthaul capacity $C$ for $a_U=0.2$. We first note that, for small values of $C$ (in our case, $C\lesssim 4\ \mathrm{bit/s/Hz}$), puncturing is preferable to treating URLLC as noise, since, as explained above, it avoids wasting the limited fronthaul resources on samples that are corrupted by URLLC interference. In this regime, puncturing provides close performance to SIC, with the added benefit of a lower complexity and power consumption at the ENs. For larger fronthaul capacities, the quantization noise tends to zero, and thus treating URLLC as noise outperforms puncturing, given that it allows the BBU to make full use of the received signals. Moreover, H-NOMA with SIC provides the largest rate. Finally, both Fig.~\ref{fig:UL_function_a_U} and Fig.~\ref{fig:UL_function_C} indicate that, with a sufficiently powerful decoder, such as SIC, the eMBB rate can be improved under H-NOMA as compared to H-OMA.
\begin{figure}[t]
	\centering
	\includegraphics[height= 6.3 cm, width= 9.5 cm]{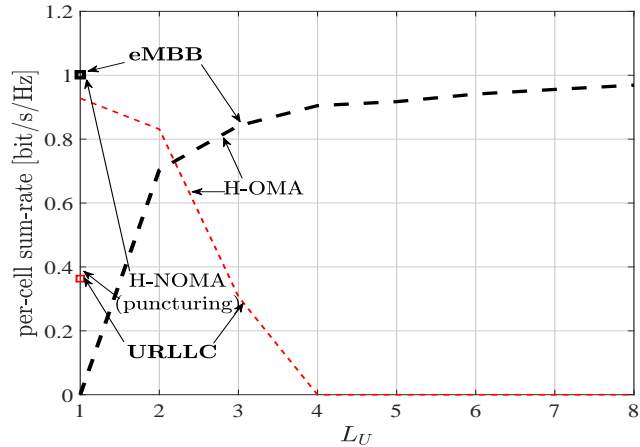}
	\caption{UL eMBB average per-cell sum-rate and URLLC per-cell sum-outage capacity as a function of the URLLC access latency $L_U$ for H-OMA and for H-NOMA with different decoding strategies ($\epsilon_U = 10^{-3},C=2\ \mathrm{bit/s/Hz}$, $\gamma=3$, $a_U=0.5 \times 10^{-3}$)}
    \vspace{-5 mm}
	\label{fig:UL_function_L_U}
\end{figure}
\par In Fig.~\ref{fig:UL_function_L_U}, we study the trade-off between the eMBB and URLLC per-user rates as a function of the access latency $L_U$. Under H-OMA, the URLLC per-user rate decreases when the access latency $L_U$ grows due to the increased probability of URLLC packet collision. In order to compensate for this contribution to the probability of error in \eqref{eq:error_oma_ul}, one needs to reduce the probability of decoding error $\epsilon_{U}^{D}$, causing the rate to decrease (see \eqref{eq:ul_oma_Ru}-\eqref{eq:ul_oma_outageproba}). In contrast to H-OMA, H-NOMA provides minimal and constant URLLC latency equal to $L_U=1$, but at the price of a lower URLLC rate due to interference from eMBB transmission.
%%%%%%%%%%%%%%%%%%%%%%%%%%%%%%%%%%%%%%%%%%%%%%%%%%%%%%%%%%%%%%%%%%%%%%%%%%%%%%%%%%%%%%%%%%%%%%%%%%%
\subsection{Downlink}
Comparison between H-OMA and H-NOMA is qualitatively similar to the UL and hence we focus here on aspects that are specific to the DL.
\begin{figure}[ht]
	\centering
	\includegraphics[height= 6.3 cm, width= 9.5 cm]{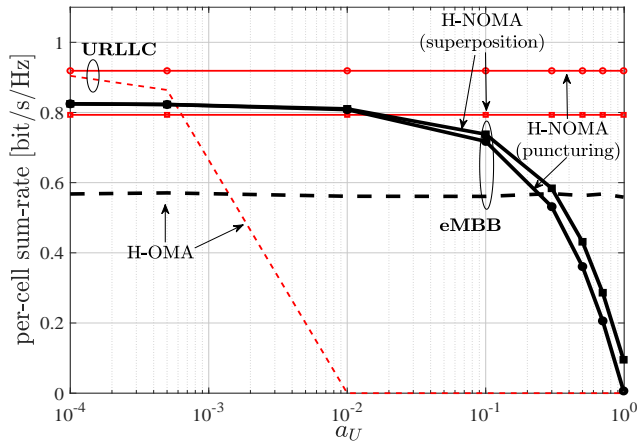}
	\caption{DL eMBB average per-cell sum-rate and URLLC per-cell sum-outage capacity as a function of the packet generation probability $a_U$ for URLLC traffic for H-OMA with $L_U=2$ and for H-NOMA with different decoding strategies ($\epsilon_U = 10^{-3},C=2\ \mathrm{bit/s/Hz}$, $\gamma=3$)}
    \vspace{-5 mm}
	\label{fig:DL_function_a_U}
\end{figure}
In Fig.~\ref{fig:DL_function_a_U}, we investigate the average eMBB per-cell sum-rate and URLLC per-cell sum-outage capacity as a function of the URLLC traffic generation probability $a_U$. As for the UL, the URLLC rate under H-OMA is seen to decrease as function of $a_U$. However, due to possibility to avoid collisions due to EN scheduling, the DL performance is limited only by blockages and hence the rate degradation is more graceful than for the UL. Another interesting aspect is the comparison between puncturing and superposition coding. Superposition coding is seen to offer a higher eMBB rate due to the larger number of mini-slots available for eMBB transmissions, but a lower URLLC performance owing to interference with eMBB signals which is absent with puncturing. Finally, unlike for the UL results in Fig. 9, H-OMA is observed to provide a larger eMBB rate than H-NOMA for larger values of $a_U$. This is because, as discussed, SIC cannot be effectively carried out in the DL. 
\par
\begin{figure}[t]
	\centering
	\includegraphics[height= 6.3 cm, width= 9.5 cm]{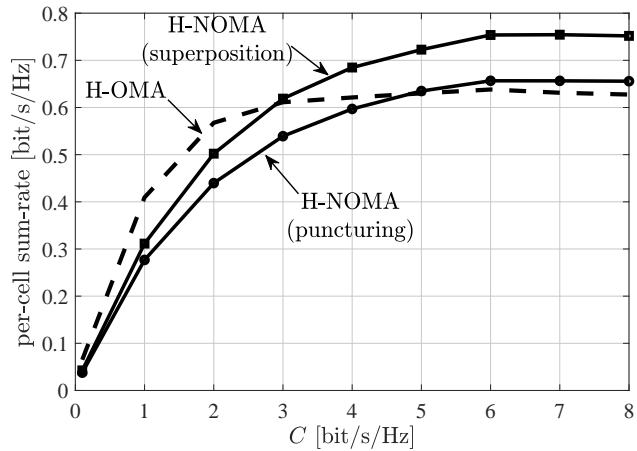}
	\caption{DL eMBB average per-cell sum-rate as a function of the fronthaul capacity $C$ for H-OMA with $L_U=2$ and for H-NOMA with different decoding strategies ($\epsilon_U = 10^{-3},\gamma=3$, $a_U=0.4$)}
    \vspace{-5 mm}
	\label{fig:DL_function_C}
\end{figure}

 We further explore the comparison between H-OMA and H-NOMA in Fig.~\ref{fig:DL_function_C}, where we plot the eMBB average per-cell sum-rate as function of the fronthaul capacity $C$ for a large URLLC packet generation probability $a_U = 0.4$. The main observation here is that, unlike the UL (cf. Fig.~\ref{fig:UL_function_C}), H-NOMA is outperformed by H-OMA for small values of $C$, here for $C \leq 3 \mathrm{bit/s/Hz}$. In fact, in the UL, H-NOMA is able to avoid using fronthaul resources for mini-slots that are affected by URLLC interference by leveraging either puncturing or SIC at the ENs. In contrast, in the DL, the BBU is unaware of the URLLC activation and hence it cannot prevent transmitting mini-slots that will eventually either be dropped at the EN, if puncturing is used, or affected by URLLC interference, if superposition coding is adopted. That said, if $C$ is large enough, H-NOMA under both puncturing and superposition coding is able to outperform H-OMA.
\section{Concluding Remarks}
\label{sec:conclusion}
This work has investigated for the first time a multi-cell F-RAN architecture in which eMBB and URLLC services may share the radio resources non-orthogonally with URLLC traffic being processed at the edge and eMBB traffic at the cloud as in a C-RAN architecture. Non-orthogonal transmission was seen to offer potentially significant gains for both eMBB and URLLC services, despite creating inter-service interference. Beside the smaller URLLC access latency, the rate gains of the resulting Heterogeneous-NOMA (H-NOMA) strategy stem from its capability to use more efficiently spectral resources for eMBB traffic, while reducing collisions and blockages for URLLC data. Nevertheless, when the URLLC activation probability is large or the fronthaul capacity is small, the advantages of H-NOMA hinge on an effective management of URLLC interference on eMBB signals. This can be done in the UL by means of puncturing or successive interference cancellation at the ENs prior to fronthaul compression. In contrast, URLLC interference management is more complex in the DL due to the lack of a priori knowledge of the central unit about URLLC activations. Among directions for future work, we mention the inclusion in the study of massive machine-type traffic.
\bibliographystyle{IEEEtran}
\bibliography{Biblio}

\end{document}